\documentclass[amsmath,comment,amssymb,twocolumn,showpacs,prb]{revtex4}
\usepackage{graphicx}
\usepackage{dcolumn}
\usepackage{bm}
\newcommand{\BEQ}{\begin{equation}}
\newcommand{\EEQ}{\end{equation}}
\newcommand{\BEA}{\begin{eqnarray}}
\newcommand{\EEA}{\end{eqnarray}}

\newcommand{\p}{\partial}
\newcommand{\nn}{\nonumber }
\newcommand{\Tr}{{\rm Tr}}

\newcommand{\ext} {\mathop{\rm ext}}

\begin{document}

\title{Marginal States in Mean Field Glasses}
\author{Markus M\"{u}ller$^\star$,
Luca Leuzzi $^{\dag,\bullet}$ and Andrea Crisanti$^\bullet$}
\affiliation{$^\star$ Department of Physics and Astronomy, Rutgers
University,
136 Frelinghuysen Road, Piscataway 08854, New Jersey, USA.
\\  $^\dag$ INFM-CNR, Via dei Taurini 19, 00185 Rome, Italy.
\\ $^\bullet$
Department of Physics, University ``La Sapienza'',
Piazzale Aldo Moro 5, 00185 Rome, Italy. }

\date{\today}

\begin{abstract}
We study mean field systems whose free energy landscape is dominated
by marginally stable states.  We review and develop various techniques
to describe such states, elucidating their physical meaning and the
interrelation between them.  In particular, we give a physical
interpretation of the two-group replica symmetry breaking scheme and
confirm it by establishing the relation to the cavity method and to
the counting of solutions of the Thouless-Anderson-Palmer
equations. We show how these methods all incorporate the presence of a
soft mode in the free energy landscape and interpret the occurring
order parameter functions in terms of correlations between the soft
mode and the local magnetizations.  The general formalism is applied
to the prototypical case of the Sherrington-Kirkpatrick-model where we
re-examine the physical properties of marginal states under a new
perspective.
\end{abstract}

\maketitle

\section{Introduction}

Disordered, frustrated systems often possess a multitude of nearly
degenerate metastable states which are at the basis of glassy
phenomena like slow relaxation to equilibrium and aging. In order to
understand the dynamics of such systems it is important to
characterize their corrugated free energy landscape, as well as the
physical properties and abundance of metastable states, usually
described by the complexity function (also called configurational
entropy in the context of amorphous solids).

In finite-dimensional, short-range interacting systems, it is not
possible to rigorously define metastable states since nucleation
phenomena always restore ergodicity on sufficiently long time
scales. However, in mean-field models where an analytical description
of the free energy landscape in terms of local order parameters
(magnetizations) is available, metastable states can be identified as
stable stationary points of the free energy functional (see
e.g. Ref. [\onlinecite{ParisiLH05}] for an instructive review of the
state of the art).  The properties of the local neighborhood of the
latter allow for a natural classification: (i) genuine minima (with a
positive definite free energy Hessian),\cite{CLRpSP} (ii) {\em
marginal states} (with eigenvalues of the Hessian tending to zero in
the thermodynamic limit). \cite{ABM,CGPnum,PRJPA04}

Recent studies of the Ising $p$-spin model at low temperature
have lead to the following picture for the metastable states as a function of their free energy density:
at low free energy densities most metastable states are genuine minima,
while above some threshold 
free energy 
entropy is dominated by marginal states with a single soft mode.\cite{CLR05,L05} The same phenomenology is found
 in mixed spherical $s+p$-spin models.\cite{Gualdi}
In the special case of
the spherical $p$-spin model,\cite{CS92,CS93} the dominant
metastable states are genuine minima at all free energy densities up to the
threshold of dynamic arrest, above which there are {\em no} stable states anymore. \cite{CLRpSP}

In certain models, however,  marginal states are exponentially more numerous
than genuine minima {\em at all free energy
densities}. This situation is expected to occur generically
in models whose thermodynamics is described by continuous replica symmetry breaking,
the best studied case being the Sherrington-Kirkpatrick (SK)
model.\cite{SKPRL75,ABM,PRJPA04}

The studies of the above models have shown that very often marginal states possess only one single soft mode.
Such states require a special treatment both in replica and cavity approaches, and we will focus on their description in the present study.
Under certain circumstances, however, marginal states with a large number (diverging
in the thermodynamic limit) of soft directions in free energy landscape may occur,
the best known example
being the thermodynamically dominant state of the SK model below the critical temperature
$T_c$.~\cite{BM79} This global free energy minimum is a stationary point
with many almost flat directions in free energy landscape.
Since there is a multitude of marginal directions
none of them can be singled out a priori. It is thus not clear whether the
 methods presented here, tailored to the presence of a single marginal mode,
 may still apply to that situation.
 \cite{fn1}

The dynamical behavior of a given model will strongly depend on the
local environment of the metastable states that dominate the landscape
in the range of dynamically accessible free energies (usually energies
where marginal state dominate).  To date, the role of higher-lying
marginal states in the slow dynamics of mean field glasses remains
unclear. However, the characterization of their local properties
presented here should help to identify their traces in future
analytical or numerical studies of glassy dynamics.

In this paper, we focus on the analytical description of marginal
states with a single soft mode and show what physical information is
contained in three choice methods to describe them: the two group
replica formalism, the generalized cavity method and the counting of
the solutions of the Thouless-Anderson-Palmer (TAP) mean-field
equations.  In particular, we show that the emerging order parameters
contain information on the correlation between the marginal mode and
the local magnetizations.  Further, we discuss the computation of the
distribution of (frozen) local fields and infer further key
characteristics of the local environment of marginal states, such as
the spectrum of the free energy Hessian.

From a detailed analysis of the SK model, it will become clear that
the compact two-group replica method \cite{BM2g,PP} is an effective
means to describe marginal metastable states in mean field systems.
We compare this approach with other methods and exhibit their
equivalence at the level of the annealed approximation.

The paper is organized as follows: in Sec.~\ref{s:class} we briefly
review various approaches to stable and marginal states and summarize
the current knowledge.  In Sec.~\ref{s:toy} we introduce an exactly
solvable toy model whose physics is dominated by marginal states.  By
analyzing it with the help of a simplified two-group Ansatz we gain an
understanding of the physical meaning of the formalism that appears in
the more complicated mean field models studied in Sec.~\ref{s:2G}.  In
Sec.~\ref{s:TAPcount} we relate the two group calculation to the
direct counting of solutions of the TAP equations.
Rederiving the TAP-complexity following Bray and Moore, we exhibit the
 equivalence with the two group approach.  In Sec.~\ref{s:cav} we
 review and extend the cavity method adapted to marginal
 states.
We show its equivalence with the two
 group formalism and confirm the interpretation of the order
 parameters.  In Sec.~\ref{s:quenched} we build the formalism for a
 {\em quenched} two-group computation and discuss its physical
 content, in particular the distribution of local fields and soft mode
 components.  In Sec.~\ref{s:stab} we recall the criteria for internal
 thermodynamic consistency and local stability, and discuss possible
 scenarios for the evolution of the local properties of metastable
 states as their free energy decreases.  In Sec.~\ref{s:lowT} we
 analyze the local field distribution in the uncorrelated, high
 energy, regime at low temperatures, and speculate on its consequences
 for the dynamics.  Finally, in Sec.~\ref{s:outl}, we discuss various
 open questions and possible future extensions of the presented
 methods, concluding with a brief summary in Sec. \ref{s:conc}.

\section{Classification of metastable states}
\label{s:class}
The choice techniques to investigate metastable states in mean
field
systems with quenched disorder are
\begin{itemize}
\item the replica method with an
ultrametric Replica Symmetry Breaking (RSB) Parisi
Ansatz,\cite{P80f}
\item  the direct counting of the
solutions of the Thouless-Anderson-Palmer (TAP)
equations,\cite{TAP,BMan}
and
\item
 the cavity method.\cite{MPV86}
\end{itemize}
 All these approaches can be used to
describe stable states and to compute their properties  as well as their
complexity, i.e., the logarithm of the
number of states at given free energy density.

However, in situations where the most numerous states are marginal,
these
techniques need to be generalized.\cite{CGP04,RizzoJPA05} As explained in the introduction,
this generally
occurs at sufficiently high free energy densities, and in certain models even
at all free energies (in the SK model below $T_c$, for instance).
 The characteristics of the three
 equivalent approaches to stable states, and their generalizations adapted to
 marginal states are summarized in table \ref{tab:I}.
\begin{table}[!h]
\begin{tabular}{|l|c|c|}
\hline
\vspace{-2mm}
\\
 & Minima & Marginal
\\
\hline
\vspace{-2mm}
\\
{\bf Replica} & Parisi RSB  & Two group+RSB
\\
& Ansatz & Ansatz
\\
\hline
\vspace{-2mm}
\\
{\bf Cavity} &  States robust &  Fragile pairs of
\\
&  to addition of a spin & saddles and minima
\\
&  Single cavity field & Additional field
\\
&  & related to soft mode
\\
\hline
\vspace{-2mm}
\\
{\bf Counting of}  & Saddle point  &  Saddle point
\\
{\bf TAP solutions}  & of TAP action &   of TAP action
\\
{\bf } & conserves   & breaks
\\
& fermionic symmetry  & fermionic symmetry
\\
\hline
\end{tabular}
\label{tab:I}
\caption{Table summarizing the methods to
study properties of metastable states in mean field glasses. The second
column refers to genuine minima, while the third column describes
the necessary generalization in the case
of {\em marginal states} with a single soft mode. }
\end{table}

\subsection{Prevalence of genuine minima}
\label{ss:genmin}
At free energy densities where the vast majority of
metastable states are
minima of the free
energy landscape any thermodynamic function (including the
complexity)
is correctly described by Parisi's RSB
 {Ansatz}, by the  standard cavity method or by
counting stable solutions of the
 TAP equations (imposing a saddle point which preserves a fermionic
symmetry of the problem\cite{ZZ,K91}).

Mean field models with an ultrametric organization of states in two
levels (a "one step" structure) usually exhibit this kind of landscape at low enough free
energies. Their static properties can be computed either using a
one
step RSB Ansatz\cite{Parisi1step}  or the cavity method including
cluster
correlations via the so-called ``reweighting''.\cite{MPV86} Whenever the
one step RSB solution is stable the complexity of
stable minima can be calculated using Monasson's method as the Legendre transform of
the free energy of $m$ coupled clones.\cite{MPRL95}

It is a non-trivial fact that the same result can be obtained by
counting the number of solutions of the TAP
equations~\cite{BMan,CLPR03} which requires the saddle point
extremization of a certain action functional. Under the assumption
that every solution represents a stable minimum, the corresponding
saddle point has to preserve the fermionic Becchi-Rouet-Stora-Tyutin
(BRST) symmetry of the action\cite{BRST} which generically occurs in the
description of stochastic equations.\cite{PLH,K91}

\subsection{Prevalence of marginal states}
In those cases where marginal states dominate (being exponentially
more numerous than genuine minima) the approaches mentioned above fail
because they assume the states to be stable minima.  However, all
approaches can be suitably generalized to deal with the marginal case,
too.  The purpose of this paper is to examine the physical meaning of
these generalizations, namely (i) the extension of the RSB scheme to a
two-group Ansatz,\cite{BM2g,PP,CLPR04b} (ii) the inclusion of an extra
auxiliary field in the cavity method \cite{RizzoJPA05,CGP04} and (iii)
the counting of TAP solutions via a saddle point that breaks the BRST
symmetry.\cite{BMan, ABM} While some aspects of the generalized cavity
method and the breaking of the BRST symmetry have been interpreted
previously in the literature, the meaning of the two group Ansatz, as
well as the relation between the above methods have not been
established. The present paper tries to fill this gap.  Furthermore,
we will show how to extract so far hidden information from the
formalism and give the interpretation of the emerging order parameter
functions.

Before applying the two group approach to the SK model, we shall
introduce a simple, exactly solvable toy model which illustrates the
basic features of the marginal states and their appearance within a
replica language. This will be a conceptual guide to the physics
discussed later in the context of the more complicated SK model.

\section{A simple model to understand the two group Ansatz and
marginal states}
\label{s:toy}
The two group Ansatz was first introduced by Bray and Moore in
1978,
Ref.  [\onlinecite{BM2g}], in an attempt to resolve the instability of
the
replica symmetric solution found by Sherrington and
Kirkpatrick\cite{SKPRL75} in the mean-field approximation of the
Edwards-Anderson model.\cite{EA75} Even though it did not
turn out to be the correct way to describe the equilibrium
(the stable equilibrium solution with ultrametric
replica symmetry breaking was found soon after by Parisi\cite{P80f}), this two group Ansatz endowed
with an ultrametric structure can be used to study the
free energy landscape {\em above} the thermodynamic ground state.

This replica symmetry breaking scheme consists in dividing $m$
replicas in two non-equivalent groups of $m+K$ (group ``$+$'') and
$-K$ (group ``$-$'') elements, respectively, and computing the
corresponding "replicated free energy".  A subsequent Legendre
transform with respect to the total number $m$ of
replicas\cite{MPRL95} was found to reproduce Bray and Moore's
calculation of the TAP complexity.~\cite{BMan,BM2gcomp,PP}

A very similar scheme of replica symmetry breaking with groups of
$K\rightarrow \pm \infty$ replicas occurred in the mean field
description of the random field Ising
model,~\cite{PLH,DotsenkoParisi93,DotsenkoMezard97} where "instantons"
with this kind of two-group structure were used to identify a certain
class of Griffith-like singularities of the free
energy,~\cite{Dotsenko06} the physics of which will become clear from
the following toy model.\cite{fn2}

\subsection{The meaning of the two group limit in a toy model}
In this section we examine an exactly solvable zero-dimensional
model which contains the basic ingredients to an understanding of the two
 group Ansatz.

From the replica solution of this toy model we will infer that the two
groups should be interpreted as representing a minimum and a saddle
point that merge into a single marginally stable configuration in the
limit $K\rightarrow \infty$. Furthermore, this picture will allow us
to obtain a physical interpretation of the order parameters appearing
in the two-group Ansatz, as we will later confirm using the
equivalence with the physically more intuitive cavity approach.

Let us consider the simple model of a particle in a potential
$V(\phi)$ and subject to a random field $h$ with probability
distribution $P(h)$,
\begin{equation}
\label{model}
{\cal H}(\phi;h)=V(\phi)-h\phi.
\end{equation}
This can be considered  as the $0$-dimensional case of the problem of
pinned manifolds, such as the Random Field or Random Temperature
Ising Model where a very similar analysis
leads to the description of Griffith
singularities.~\cite{Dotsenko06}

We consider potentials such that for typical fields $h$ the
Hamiltonian ${\cal H}$ possesses only one minimum at $\phi_I(h)$. A
secondary
minimum occurs at $\phi_{I\!I}(h)$ only for rare fields that are larger
than a critical value $h>h_c$. A simple example for such a
potential
is given by
\begin{equation}
\label{potential}
V(\phi)=h_c
(\phi-\phi_c)+\frac{V_3}{6}(\phi-\phi_c)^3,
\end{equation}
subject to the constraint $\phi\geq 0$.
In this case, we have
$\phi_I(h)\equiv 0$ for all random fields of practical interest, and for
$h\geq h_c$ there is a secondary minimum $\phi_{I\!I}(h)$. The latter
becomes marginal as $h\rightarrow h_c$, while $\phi_{I\!I}$ approaches
$\phi_{I\!I}(h_c)=\phi_c$.

\begin{figure}[tbp]
\resizebox{0.4\textwidth}{!}{\includegraphics{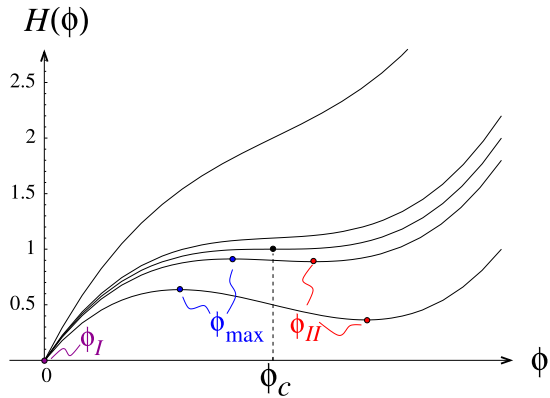}}
\caption{Hamiltonian $H(\phi)=V(\phi)-h\phi$ with $V(\phi)$ from
Eq.~(\ref{potential}), plotted for $h/h_c =0,0.9,1,1.1$ and $1.5$
(from top to bottom). The secondary minimum $\phi_{I\!I}$ and the
saddle $\phi_{\rm max}$ merge for $h\rightarrow h_c$.}
\label{figpot}
\end{figure}

 At sufficiently low temperature ($T\ll V(\phi_c)-V(0)$) the
 disorder average of the free energy behaves as
\begin{eqnarray}
\overline{F}&\equiv& -\beta^{-1} \overline{\ln Z} =
-\beta^{-1}\int dh P(h) \ln\left[\int d\phi~ e^{-\beta {\cal
H}(\phi,h)}\right]
\nonumber\\
& \approx& V(0)+\int_{h_c}^{\infty} dh\, P(h)
e^{-\beta\left[{\cal H}(\phi_{I\!I}(h);h)-{\cal H}(\phi_{I}(h);h)\right]},
\label{freeenergy}
\end{eqnarray}
where we expanded the logarithm and made a saddle point approximation
to obtain the second term describing the Griffith contribution from
the secondary minimum in rare random fields.

For simplicity, we will carry out explicit calculations for the case
of Gaussian random fields with distribution
\begin{equation}
\label{GaussPhtoy}
P(h)=\frac{1}{\sqrt{2\pi h_0^2}}\exp\left[-\frac{h^2}{2h_0^2}\right].
\end{equation}
In this case, the integral in Eq.~(\ref{freeenergy}) is dominated by
its lower boundary for $T\gtrsim T^*\equiv \phi_c h_0^2 /h_c$,
\begin{eqnarray}
\label{freeenergy2}
\overline{F}\approx V(0)+{\rm const.} \times P(h_c) e^{-\beta\Delta E_c},
\end{eqnarray}
where $\Delta E_c$ is the energy difference between primary and
secondary minimum, evaluated at the critical field strength, $h_c$,
\begin{eqnarray}
\label{DeltaF}
\Delta E_c=V(\phi_c)-V(\phi_I) -h_c(\phi_c-\phi_I).
\end{eqnarray}

\subsubsection{Derivation with vectorial replica symmetry breaking}

It is instructive to rederive this simple result with a replica
calculation. Dotsenko and M{\'e}zard~\cite{DotsenkoMezard97} found
that the exact low temperature partition function of similar
disordered single particle models could be reproduced by summing a
class
 of saddle points that break replica symmetry in a non-standard
(i.e., non-ultrametric), "vectorial" way. The physical content of this
recipe will become clearer below, where we will show that
a generalization of their scheme applied to the model
EQ. (\ref{model}) indeed gives back the anticipated
result of Eq.~(\ref{freeenergy2}).

For a Gaussian field distribution, Eq.
(\ref{GaussPhtoy}), the replicated and averaged partition function reads
\begin{eqnarray}
\label{Zn}
\overline{Z^n}&=&\int \prod_{a=1}^n d\phi_a \exp[-\beta
F(\{\phi_a\})] \\
&=&\hspace*{-.2cm} \int \! \prod_{a=1}^n d\phi_a \exp\left[ -\beta \sum_{a=1}^n
V(\phi_a)
+\frac{h_0^2}{2}\left(\beta \sum_{a=1}^n\phi_a\right)^2
\right].\nonumber
\end{eqnarray}
For arbitrary distributions $P(h)$,
 characterized by their cumulants $c_r\equiv \langle h^r\rangle_c$,
the replica free energy functional generalizes to
\begin{equation}
\label{Fn}
F(\{\phi_a\})=\sum_{a=1}^n
V(\phi_a)-\sum_{r=1}^{\infty}\frac{1}{\beta}\frac{c_r}{r!}\left(\beta
\sum_{a=1}^n\phi_a\right)^r.
\end{equation}

Following the recipe by Dotsenko and
M\'ezard~\cite{DotsenkoMezard97},
we approximate the integral in Eq.~(\ref{Zn}) by determining all
stable
saddle points of the free energy functional $F(\{\phi_a\})$, and
summing their respective Boltzmann weights,
\begin{eqnarray}
-\beta \overline{F}\equiv \overline{\ln Z}=\lim_{n\rightarrow
0}\frac{\overline{Z^n}-1}{n}\approx \frac{Z_{\rm
RS}-1}{n}+\frac{Z_{\rm RSB}}{n}.
\end{eqnarray}
The partition function
$Z_{\rm RS}=\exp[-\beta F_{\rm RS}]$ denotes the contribution of
the
replica symmetric saddle point with $F_{RS}\equiv
F(\{\phi_a=\phi_{\rm
RS}\})= nf_{\rm RS}$, and $Z_{\rm RSB}$ is the sum over all
saddle points breaking the replica symmetry. As we will see
shortly, the latter always comes with a combinatorial factor
proportional to $n$, which cancels the denominator.

Let us show that this recipe allows us to rederive Eq.
(\ref{freeenergy2}).  We make the most general Ansatz for the
configuration $\{\phi_a\}$ of a saddle point, collecting the
$\phi$'s
with identical values into $M$ groups labeled by $i=1,\dots,M$,
each
with $k_i$ replicas, i.e.,
\begin{equation}
\label{Ansatz}
\begin{array}{lcl@{\quad\quad}l}
 \phi_a&=&\phi_1, & {a=1,\dots,k_1,}\\
 \noalign{\vskip 5 pt}
 \phi_a&=&\phi_2, & {a=k_1+1,\dots,k_1+k_2,}\nonumber\\
 \noalign{\vskip 5 pt}
 &\vdots& & {}\nonumber\\
 \noalign{\vskip 5 pt}
 \phi_a&=&\phi_M, & {a=(\sum_{i=1}^{M-1}k_i)+1,\dots,n.}\nonumber\\
  \end{array}
\end{equation}

In the Gaussian case, the corresponding replica free energy
evaluates to
\begin{equation}
\label{Frep}
F_{\{k_i\}}=\sum_{i=1}^M k_iV(\phi_i)-\frac{\beta h_0^2}{2}
\left( \sum_{i=1}^M k_i\phi_i \right)^2,
\end{equation}

The saddle point equations with respect to variation of $\phi_i$
read
\begin{eqnarray}
\label{SPrep}
&&V^{\prime}(\phi_i)=h_{\phi}, \hspace*{2cm} \forall i,
\\
\label{toySC}
&&h_{\phi}\equiv \beta h_0^2 \sum_{i=1}^M k_i\phi_i.
\end{eqnarray}

For a regular potential $V(\phi)$, the number $M$ of different
solutions of Eq.~(\ref{SPrep}) is limited. In particular, for the
potential in Eq.~(\ref{potential}) we have $M=3$ for $h_\phi>h_c$,
$M=2$ for $h_\phi=h_c$, and $M=1$ for $h_\phi<h_c$. The latter
corresponds to the replica symmetric saddle point with $h_\phi=0$,
$\phi_a=0$, $F_{\rm RS}= V(0)$.

Replica symmetry breaking saddle points exist for $h_\phi\geq h_c$,
where we label the global minimum, the secondary minimum and the local
maximum of ${\cal H}$ by $\phi_I$, $\phi_{I\!I}$ and $\phi_{\rm max}$,
respectively, leaving their dependence on $h$ implicit
(cf. Fig.~\ref{figpot}).  In the limit of $n\rightarrow 0$,
Eq.~(\ref{toySC}) yields the self-consistency equation for $h_\phi$
\begin{equation}
\label{selfconsistency}
k_1(\phi_I-\phi_{\rm max})+k_2(\phi_{I\!I}-\phi_{\rm max})=\frac{h_\phi}{\beta
h_0^2}.
\end{equation}

Notice, that we choose the number of
replicas in the minima, $k_1$ and $k_2$, to be positive, leaving a
negative number $n-k_1-k_2\rightarrow -(k_1+k_2)$ of replicas in
$\phi_{\rm max}$. This is necessary to ensure a positive Hessian of the
free
energy functional Eq.~(\ref{Fn}).

Let us now consider the saddle point free energy $F_{k_1,k_2}$, Eq.
(\ref{Frep}), as a function of $k_2$ for fixed $k_1$. For not too low
temperatures $T>T^*$, $F_{k_1,k_2}$ decreases as $k_2$
increases. Hence, the most important contribution to the sum over
saddles comes from large $k_2 \gg 1$. As is evident from
Eq.~(\ref{selfconsistency}), this requires $\phi_{I\!I}$ and
$\phi_{\rm max}$ to approach each other, and hence, $h_\phi$ must be
nearly critical.  More precisely, one finds
\BEA
\phi_{I\!I}&\approx& \phi_c+ \frac{h_c}{2\beta h_0^2}\frac{1}{k_2}
+\dots
,\\
\phi_{\rm max}&\approx& \phi_c- \frac{h_c}{2\beta h_0^2}\frac{1}{k_2}
+\dots
,\\
h_\phi&\approx&
h_c+\frac{V_3}{2}\left(\frac{h_c}{2\beta
h_0^2}\right)^2\frac{1}{k_2^2}
+\dots
,\\
\label{Fk1k2}
F_{k_1,k_2}&=&
k_1 \Delta E_c+\frac{h_c^2}{2\beta
h_0^2}+O\left(\frac{1}{k_2^2}\right).
\EEA

The total Griffith contribution to the free energy results from the
sum over all RSB saddle points with the corresponding multiplicity,
\begin{eqnarray}
\label{saddlesum}
-\beta F_{\rm RSB}&=&\lim_{n\rightarrow 0} \frac{Z_{\rm RSB}}{n}\\
&=&\lim_{n\rightarrow 0}\frac{1}{n}\sum_{k_1>0,k_2\geq 0}
\left( \begin{array}{c}
 n\\
k_1,k_2
  \end{array}\right) e^{-\beta F_{k_1,k_2}}\nonumber\\
& =&\hspace*{-.3cm}\sum_{k_1>0,k_2\geq 0}\hspace*{-.3cm}
\frac{(-1)^{k_1+k_2-1}(k_1+k_2-1)!}{k_1!k_2!}  e^{-\beta
F_{k_1,k_2}}.\nonumber
\end{eqnarray}
In order to recover Eq.~(\ref{freeenergy2}), we need to exclude
saddles with $k_1=0$. This can be understood on physical grounds:
the saddle point free energies $F_{k_1=0,k_2}$ are independent of the
ground state level $\phi_I$. The corresponding terms are not
suppressed by the Boltzmann weight associated with the excitation
probability to the state $\phi_{I\!I}$. In other words, these terms don't know
 about the ground state and hence must be discarded for the calculation of
the Griffith correction.\cite{fn3}

We note that for any fixed
$k_1>0$, the sum over $k_2$ is weakly divergent and has to be
regularized appropriately. Since the sum is dominated by large
$k_2$, we use Eq.~(\ref{Fk1k2}) to
approximate the Griffith contribution

\begin{eqnarray}
-\beta F_{\rm RSB}\!&\approx&\! \left\{\sum_{K> 0} \frac{(-1)^{K-1}}{K}
\sum_{k_2=0}^{K}
\left( \begin{array}{c}
 K\\
k_2
  \end{array}\right) e^{-\beta (K-k_2) \Delta E_c}\right.
\nonumber
\\
 &&\hspace*{2.7 cm}\left.-\sum_{k_2>0} \frac{(-1)^{k_2-1}}{k_2}\right\}
e^{-\frac{h_c^2}{2 h_0^2}}
\nonumber
\\
&&\hspace*{-1.1 cm}
=e^{-\frac{h_c^2}{2 h_0^2}} \left\{
\sum_{K>0}\frac{(-1)^{K-1}}{K}
\left(1+e^{-\beta \Delta E_c}\right)^K - \ln 2\right\}
\nonumber
\\
&&\hspace*{-1.1 cm}
=e^{-\frac{h_c^2}{2 h_0^2}}\ln\left(1+\frac{e^{-\beta \Delta
E_c}}{2}\right)
 \approx \frac{1}{2}e^{-\frac{h_c^2}{2 h_0^2}}e^{-\beta \Delta E_c}
\nonumber
\\
&&\hspace*{-1.1 cm}
\approx P(h_c) e^{-\beta \Delta E_c}.
\label{toy:f_rsb}
\end{eqnarray}
and we recover indeed the Griffith correction, Eq.
(\ref{freeenergy2}), due to
rare secondary minima.

An analogous analysis can be carried out in the case of
non-Gaussian
distributions $P(h)$. Instead of Eq.~(\ref{Fk1k2}), one obtains the
saddle point free energy
\begin{eqnarray}
\label{Fk1k2general}
F_{k_1,k_2\rightarrow \infty}&=&k_1 \Delta
E_c-\beta^{-1}S_c+O(1/k_2^{2}).
\end{eqnarray}
where $S_c \approx \log[P(h_c)]$. More precisely, $\exp(S_c)$ is
given
by the saddle point approximation of the integral representation
\BEA
P(h_c)&=&\int \frac{d \lambda}{2\pi} \exp\left(-i\lambda
h_c+\sum_{r=1}^{\infty}\frac{c_r}{r!}
(i\lambda)^r\right)\nonumber\\
&\approx&\ext_{\lambda_*}\left[\exp\left(-\lambda_*
h_c+\sum_{r=1}^{\infty}\frac{c_r}{r!} \lambda_*^r\right)\right],
\EEA
that is a good approximation provided that $h_c$ belongs to the
tail
of $P(h)$.  The Griffith term is again dominated by $k_1=1$ and
$k_2\rightarrow \infty$, yielding $\Delta F_c \approx P(h_c)
\exp[-\beta \Delta E_c]$.

In summary, the above toy model describes a simple Griffith phase
which is dominated by marginal configurations.
This physics is exactly reproduced by a vectorial
replica symmetry breaking that divides the replicas into two
(infinite) groups describing
a coalescing pair of a minimum and a saddle point ($\phi_{I\!I}$ and
$\phi_{\rm max}$).

\subsection{Generalization to higher dimensions}
\label{ss:d_toy}
To make contact with more complicated models, it is instructive to
generalize our simple model to a particle in $d$ dimensions,
$\vec{\phi}\in \mathbb{R}^d$, subject to Gaussian random fields
\begin{equation}
\label{HRF}
{\cal H}(\vec{\phi};\vec{h})=V({\vec \phi})-{\vec h}\cdot {\vec
\phi}.
\end{equation}

The minima of (\ref{HRF}) satisfy ${\vec \nabla} V({\vec \phi})={\vec
h}$. The Griffith contributions to the quenched free energy are
dominated by rare fields $\vec{h}_c$ that are just strong enough to
admit a marginal state $\vec{\phi}_c=\vec{\phi}(\vec{h}_c)$. The
marginality implies the presence of a soft mode in the Hessian of
${\cal H}$, i.e.,
\begin{equation}
\label{marginalityD}
\det\left[{\rm Hess}(\vec{\phi}_c)\right]\equiv
\det\left[\frac{\partial^2V(\vec{\phi}_c)}
{\partial \phi_i\partial\phi_j}\right]=0.
\end{equation}
This condition defines a hypersurface ${\cal S}$ in the space of
fields $\vec{h}$. The dominating Griffith contribution derives from
$\vec{h}_c\in {\cal S}$ which maximizes $P(\vec{h}_c)\exp\{-\beta
[{\cal H}(\vec{\phi}_c)-{\cal H}(\vec{\phi}_I)]\}$.
 In the replica formalism this
condition is conveniently encoded in the saddle point equation for
$k_1=1$ and $k_2\rightarrow \infty$, analogous to
Eq.~(\ref{selfconsistency}),
\begin{equation}
\label{selfconsistencyhighD}
(\vec{\phi}_I-\vec{\phi}_{c})+\vec{\psi}=-\frac{1}{\beta}\vec{\nabla}
\log[P(\vec{h}_c)]=\frac{\vec{h}_c}{\beta h_0^2},
\end{equation}
where
\BEQ
\vec{\psi}=\lim_{k_2\rightarrow\infty} k_2
[\vec{\phi}_{I\!I}-\vec{\phi}_{\rm max}]
\EEQ
 is proportional to the soft mode of the Hessian (computed in
$\vec{\phi}_c$), and $\vec{\phi}_I$
is the primary minimum in the presence of the field $\vec{h}_c$.

\subsection{Stabilizing marginal states}
Marginal states are rather delicate to work with since they are very
sensitive to perturbations. A way to circumvent this problem is to
introduce a regularizer favoring (slightly) stable states, and let it
tend to zero at the end of the calculation. Such an approach has been
implemented in Refs. [\onlinecite{RizzoJPA05, PRPRB05}] for the SK model
and the Viana-Bray model.  Here, we show explicitly the mechanism of
this procedure for the toy model studied above. A similar analysis for
the SK model will be presented in Sec.~\ref{s:cav} (see also
App.~\ref{app:regularization}).

Let us reconsider the one-dimensional model, Eq.~(\ref{model}). For
slightly supercritical random fields, $h=h_c+\delta h$, there is a
nearly marginal secondary minimum $\phi_{I\!I}\approx \phi_c+(2\delta
h/V_3)^{1/2}$.
In order to lift the marginality of the dominating states, we impose
a
stabilizing ``constraint'' by introducing a weight factor
$\exp[\lambda_X X(\phi)]$ in the phase space average of the free
energy. The function $X(\phi)$ is arbitrary up to the conditions
$X(0)=0$ (in order not to couple to the ground state) and
$X'(\phi_c)\neq 0$. Any other specifics of this regularizer will disappear
in the end.
The Griffith part of the quenched free energy Eq.~(\ref{freeenergy})
then
reads
\begin{equation}
\label{constraint}
F_G=\int_{h_c}^{\infty} dh\, P(h)
e^{-\beta\left[{\cal H}(\phi_{I\!I};h)-{\cal H}(\phi_{I};h)\right]
+\lambda_X X(\phi_2)}.
\end{equation}
The integrand is maximal at the saddle point $h^*=h_c+\delta h$
with
\begin{equation}
\delta h^{1/2}\approx
-\frac{\lambda_X X'(\phi_c)}{(2V_3)^{1/2}}
\frac{P(h_c)}{P'(h_c)},
\end{equation}
corresponding to a state $\phi_{II}$ which is a genuine local minimum
with a minimal eigenvalue of the energy Hessian which is positive
\BEA
\lambda_{\rm min}&=&V''(\phi(h^*))
\\
\nn
&=&\sqrt{\frac{V_3~ \delta h}{2}} \sim \lambda_X.
\EEA
The linear response to a Hamiltonian perturbation ${\cal
H}\rightarrow
{\cal H}-h_X X(\phi)$ diverges as $\partial \phi/\partial h_X\sim
\lambda_X^{-1}$ as the constraint is lifted. However, the product
\BEA
z\equiv \lim_{\lambda_X\to 0}
\lambda_X \frac{\partial \phi}{\partial h_X} =
-\frac{P(h_c)}{P'(h_c)}
\EEA
tends to a finite limit.

The above reasoning holds for any function $X$ provided
 that $X'(\phi_c)\neq 0$ which is necessary to lift the
 marginality. In the higher dimensional case, this generalizes to the requirement
\BEQ
\vec{\nabla}_\phi X\left(\vec{\phi}_c\right) \cdot \vec{\psi} \neq
0
\label{conditionTOY}
\EEQ
 where
$\vec{\psi}$ is the soft mode of the Hessian.

As in the one-dimensional case, there is a finite limit for the
combination
\BEQ
\lim_{\lambda_X\to 0}
\lambda_X \frac{\partial \vec{\phi}}{\partial h_X} =
\left(\vec{\nabla} \log[P(h_c)] \cdot \vec{\psi}\right)
\vec{\psi}.
\EEQ

The avoidance of marginality by means of a control
parameter coupled to a generic weight function $X$ will be used again in
Sec. \ref{s:cav} where we will apply the same trick to the SK
model.

\subsection{Discussion and connection to the two group
Ansatz for mean field models}

The vectorial replica analysis presented in this section
demonstrates
that sending the number of replicas of two separate groups to plus
and
minus infinity, respectively, encodes the marginality of the metastable states
 they describe.  In particular, we observe that
the sum over RSB saddles in Eq.~(\ref{toy:f_rsb}) is dominated by
terms with $k_1=1$ and $k_2\rightarrow \infty$. The latter
reflects
the fact that the leading (Griffith-like) contribution to the free
energy is due to field realizations in which the secondary minimum
is
just marginal.

Furthermore, in the $d$-dimensional case, we have seen that the
difference vector
$\vec{\psi}$ between the secondary minimum
${\vec{\phi}_{I\!I}}$ and the saddle ${\vec{\phi}_{\rm max}}$,
corresponds to the direction of the soft mode in that marginal state. This
is shown to emerge (in Sec. \ref{ss:d_toy}) assuming that the integral
over the field $\vec{h}$ is dominated by a small vicinity of a
single
saddle point $\vec{h}_c$ where $P(\vec{h})$ takes its maximum on
the
marginal surface $\cal S$ defined by Eq.~(\ref{marginalityD}).

The contribution to the partition function is dominated by a single
state only as long as the number of minima does not grow exponentially
with increasing disorder strength. This assumption breaks down in high
dimensional mean field models such as the SK model where the
dimensionality grows with number of spins $d \sim N$. Nevertheless, in
these cases the most numerous states in a given quenched realization
of random couplings turn out to be marginal, even though the origin of
marginality is different.  For any disorder configuration, there are
exponentially many states (solutions of the TAP equations) in
magnetization space, but stable states are usually less abundant than
marginal ones since true stability imposes extra constraints on the
TAP solutions. This argument certainly holds if no constraint is
imposed on the free energy density. However, in many models (e.g.,
Ising $p$-spin models), there is a range of low free energies where
typical states are stable, while at higher energies the majority of
states are marginal. The SK model is special in that it does not
possess such a low energy regime, so that the dominant states at all
free energies are marginal.

The above mechanism for marginality in mean field glasses is to be
contrasted to the toy model where marginality is a consequence of
maximizing Griffith contributions over the rare disorder. Despite this
difference, the insight from the toy model will help us in the next
section to obtain a deeper understanding of the physical significance
of the two group Ansatz for mean field glasses. In later sections we
will confirm this interpretation by revisiting the counting of TAP
states and the cavity approach.

\section{The Two Group Ansatz for mean field spin glasses}
\label{s:2G}
\subsection{Marginal states in the SK model }

Bray and Moore~\cite{BM2gcomp} discovered that
their computation of the TAP-complexity\cite{BMan} could be exactly
reproduced
by substituting for each entry in a standard
Parisi matrix a two-group matrix (with $m+K$ and $-K$ replicas, respectively),
and Legendre transforming the result with respect to $m$, following Monasson's approach.
\cite{MPRL95} Later, Parisi and Potters~\cite{PP} showed that this equivalence
extends to the low energy regime of the SK model, where full replica symmetry breaking occurs,
and also holds in a model of random orthogonal matrices.~\cite{PP2}
However, the meaning of this remarkable result remained unclear.

The preceding analysis of our toy model suggests to interpret the two
group Ansatz as a means to force representatives of the two groups
into pairs of almost coalescing minima and saddles. This picture is
strongly supported by the recent analytical \cite{ABM} and numerical
\cite{CGPnum,ABBM} finding that solutions of the TAP equations always
appear in pairs, one being a local minimum and the other a saddle of
rank one.  The straight connection of such a pair of stationary points
defines a path in the energy landscape that is increasingly flat with
increasing system size.  In the thermodynamic limit, the Hessian
matrix computed at the minimum has a zero eigenvalue with a soft mode
pointing in the direction of the adjacent saddle.

The only TAP-state without a 'partner state' is the paramagnetic
TAP solution $\{m_i=0\}$ which has to be discarded because it is unphysical.
The absence of a 'trivial' ground state constitutes an (inessential) difference between the SK model and the toy model
considered in the previous section: in the SK model
there is no physically relevant analog of the
global minimum $\phi_1$ of Sec.~\ref{s:toy}.
The marginal states do not merely occur as high energy excitations above
some trivial ground state, but they are {\em the} dominant metastable
states at a given free energy.  Therefore, a
vectorial replica symmetry breaking with only two groups of
replicas
(associated to minimum and saddle) captures all the information we
need.

\subsection{The order parameters of the two group Ansatz}
\label{ss:2gop}
In
the presence of pairs of minima and saddles, the
concept of overlaps between different
states (similarity of their magnetizations) needs to be generalized
to cover three cases: overlaps between two minima, between two saddles,
and between a minimum and a saddle. They can be described by matrices $Q^{++}_{ab}$, $Q^{--}_{ab}$,
 and $Q^{+-}_{ab}$, respectively. The replica indices $a,b$
indicate the distance within an ultrametric Parisi tree of the
respective pairs.

Such order parameters indeed appear in the two group Ansatz,~\cite{CLPR04b}
where one works with two groups of $m+K$ and $-K$ replicas.
$K$ plays now the same role as $k_2$ for the toy model of
Sec.~\ref{s:toy}. We thus expect that a set of $m$ replicas corresponds to
 a pair of a minimum and a saddle whose magnetizations differ by a quantity of the order of $O(1/K)$,
\begin{equation}
\label{2G:pheno}
  m^{\pm}_i=\overline{m}_i\pm \frac{\delta m_i}{2K},
\end{equation}
where $\{\overline{m}_i\}$ is the set of average site-magnetizations of
the
saddle-minimum pair and $\{\delta m_i\}$ is the $i$'th component of the soft mode
connecting the minimum to the nearby saddle in configuration space.
Consequently, we expect the overlap matrices to be given by
\begin{eqnarray}
\label{Qpm}
Q^{+-}_{ab}&\equiv&Q_{ab}= \overline{Q}_{ab} -\frac{C_{ab}}{4K^2},\\
Q^{++}_{ab}
&=& \overline{Q}_{ab}+\frac{A_{ab}}{K} +\frac{C_{ab}}{4K^2}
\label{Qpp}
\\
\nn
&=&Q_{ab}+ \frac{A_{ab}}{K} +\frac{C_{ab}}{2K^2},
\\
\label{Qmm}
Q^{--}_{ab}
&=& \overline{Q}_{ab}-\frac{A_{ab}}{K} +\frac{C_{ab}}{4K^2}
\\
&=&Q_{ab}- \frac{A_{ab}}{K} +\frac{C_{ab}}{2K^2},
\nn
\end{eqnarray}
with the following interpretation of the order parameters
\begin{eqnarray}
\label{QOP}
\overline{Q}_{ab}&=&
\frac{1}{N}\sum_{i=1}^N \overline{m}^a_i\overline{m}^b_i,\\
\label{AOP}
A_{ab}&=&\frac{1}{N}\sum_{i=1}^N \overline{m}^a_i\delta m^b_i,\\
\label{COP}
C_{ab}&=&\frac{1}{N}\sum_{i=1}^N \delta m^a_i\delta m^b_i,
\end{eqnarray}
that are assumed to be self-averaging and only dependent on the
distance between the minimum-saddle pairs labeled by $a$ and $b$. Note
that in particular, $Q\equiv\overline{Q}_{aa},A\equiv A_{aa}$ and
$C\equiv C_{aa}$ describe the internal overlaps of a single marginal
minimum-saddle pair.

Here we have introduced $Q_{ab}\equiv Q_{ab}^{+-}$ to match the
notation
of Ref.~\onlinecite{CLPR04b}, but the difference between $Q_{ab}$
and
$\overline{Q}_{ab}$ of order $1/K$ is immaterial in the two group
limit where $K$ is sent to infinity.

In a situation where the dominant states are stable
the matrix $\overline{Q}$ is the only order parameter
(formally there is no soft
mode,
${\delta\vec {m }}\equiv 0$ and thus, $A_{ab}=C_{ab}=0$).   In the
case of
marginal states, the matrix $A_{ab}$  measures the
correlations between the magnetization of a state ($\{\overline{m}_i^a\}$)
and the direction of the soft mode of another
state ($\{\delta m_i^b\}$) at a phase space distance labeled by
$a$ and $b$ (and vice versa).
The matrix $C_{ab}$ measures the  similarity between the
soft modes of such states.  Note that the picture of merging minima and saddles gives
 a clear interpretation only to the {\em direction} of the soft mode, while
it does not determine the magnitude of $\delta m_i$.
In order to extract physical information we should thus normalize the soft
 modes
by $\delta\hat{ m}_i=\delta m_i/\sqrt{\langle \delta m_i^2\rangle} =\delta m_i/\sqrt{C_{aa}}$.  In particular, we may
infer that $C_{ab}/C_{aa}$ describes the decreasing correlations
between soft modes with increasing distance of pairs in phase
space.
Moreover, the angle $\gamma_{ab}$ between the magnetization vector of and the soft mode in states labeled by $a, b$ is given by
\begin{eqnarray}
\cos(\gamma_{ab})&=&\frac{\langle \delta m_i^a\,m_i^b\rangle}{\sqrt{\langle \delta m_i^2\rangle\langle  m_i^2\rangle}}
=\frac{A_{ab}}{\sqrt{Q_{aa} C_{aa}}}.
\label{f:gamma}
\end{eqnarray}

If we assume that the slow relaxation dynamics follows essentially
the soft mode of marginal states, we expect that the larger
 $\gamma_{aa}$ the smaller the relative decrease of the self-overlap in the
 course of the dynamics.

\subsection{SK model: The replicated free energy}

Let us now turn explicitly to the SK model with
 the Hamiltonian
\BEA
\label{f:sk_ham}
{\cal H}=-\sum_{i<j } s_i J_{ij} s_j,
\EEA
where $J_{ij}$ is a random Gaussian matrix of zero mean and variance
$1/N$, coupling all $N$ Ising spins $s_i$ together.

Using Monasson's clone method to access higher-lying
metastable states, one computes the quenched free energy of $m$
copies
of the system,
\BEQ -\beta
F_m= \lim_{n\to 0}\frac{1}{n}\log{\overline{ Z^{mn}_J}}
=\ext_f\left[
{\overline{\log {\cal{N}}_J(f)}}- \beta m N f\right].
\label{logZmn}
\EEQ
$Z^{mn}_J$ is the partition function of $n\times m$ copies
of
a system with Hamiltonian (\ref{f:sk_ham}). $n$ is the number of replicas introduced to compute the quenched
average, while $m$ is the number of real copies (the
Legendre
conjugate of the free energy density, see Eq.~(\ref{LT}) below). In order to capture marginal states, we divide the $m$ replicas
further into two groups of $m+K$ and $-K$ elements each. As in the toy model, $K$ is
sent to infinity in the end, which forces the two groups (at fixed
replica index $a \in \{1,\dots,n\}$) into marginal minimum-saddle
pairs.

In the thermodynamic
limit $N\rightarrow \infty$, the left hand side of Eq.~(\ref{logZmn}) yields the two group
replicated
free energy
\BEQ -\beta m \Phi_{\rm 2G}(m)
\equiv\lim_{N\to\infty}\frac{-\beta F_m}{N}=
\lim_{n\to 0}\lim_{N\to\infty}\frac{1}{n N}\log{\overline{
Z^{mn}_J}},
\label{Phi2G}
\EEQ
while the right hand side evaluates to
\BEQ
\label{LT}
\frac{1}{N}{\overline{\log
{\cal{N}}_J(f)}}- \beta m  f \equiv \Sigma(f)-\beta m f,
\EEQ
where as usual we assume an exponential growth of the number of
metastable states with system size,
${\cal{N}}_J(f)\sim\exp[N\Sigma(f)]$, the latter defining the
complexity function $\Sigma(f)$.  From Eq. (\ref{LT}) the function $\beta
m \Phi_{\rm 2G}(m)$ is thus seen to be the Legendre transform of the
complexity
with respect to the pair of conjugated variables $f$ and $\beta m$.

After a standard Hubbard-Stratonovich decoupling,
one obtains the averaged partition function in terms of a functional integral over a $n m\times
n m$
replica coupling
matrix ${\cal Q}$,
\BEA
\label{Eq_Zmn}
&&{\overline{ Z^{mn}_J}}=
\int {\cal D Q}
\exp\left\{nN{\cal F}[{\cal Q}]\right\},
\\
\label{Eq_ZMN}
&&{\cal F}[{\cal Q}] =\frac{1}{n} \log\sum_{\{s_a^i\}}
\exp\left(\frac{\beta^2}{2}\sum_{ab}\sum_{ij}s_a^i
{\cal Q}_{ab}^{ij}
s_b^j\right)\\
&&\hspace{1 cm}+\frac{\beta^2}{4}
\left(m-\frac{1}{n}\Tr {\cal Q}^2\right),
\nn
\EEA
where indices run through $a, b = 1,\ldots, n$ and $i, j= 1,\ldots,
m$,
respectively.

As motivated above, the two group Ansatz consists in writing the
matrix ${\cal Q}_{ab}^{ij}$ as $n^2$ sub-matrices
$\mathbf{Q}_{ab}$,
each of dimension
$m\times m$ of the form:
\BEQ
\mathbf{Q}_{ab}=\overbrace{\left( \begin{array}{l}Q^{++}_{ab}  \\
                                         Q^{+-}_{ab}
                           \end{array}
                           \right.}^{m+K}
       \overbrace{\left. \begin{array}{r} Q^{+-}_{ab}  \\
                                          Q^{--}_{ab}
                           \end{array}
                           \right)}^{-K}\quad ,
\label{Q2g}
\EEQ
and we adopt the parametrization (\ref{Qpm},\ref{Qpp}) for the
three sectors.
The diagonal elements are ${\cal Q}_{aa}^{ii}\equiv 0$.

The partition function (\ref{Eq_Zmn}) can be evaluated through a
saddle point computation
which leads to the thermodynamic potential
\begin{eqnarray}
\label{Eq_Phi}
-\beta m \Phi_{\rm 2G}
&\equiv&
\lim_{n\to 0}\ext_{\cal Q} {\cal F}[{\cal Q}]
\\
\nn
&=&
\lim_{n \to 0} \Biggl[\beta \phi_m+ \frac{\beta^2}{4} m(1-Q)^2
-\beta^2 A(1-Q)
\\
\nn
&&\hspace*{.5cm}
-\frac{\beta^2}{n}\sum_{ab} \Biggl(\frac{1}{2}
\left(A_{ab}^2 +Q_{ab}C_{ab}\right)
\\
&&\hspace*{2.2cm}
\nn
+mQ_{ab}A_{ab}+\frac{m^2}{4}Q_{ab}^2\Biggr)\Biggr]
\end{eqnarray}
with
\BEQ
\beta \phi_m \equiv \frac{1}{n}\log\sum_{\{s_a^i\}}
\exp\left(\frac{\beta^2}{2}\sum_{ab}\sum_{ij}s_a^i
{\cal Q}_{ab}^{ij}
s_b^j\right).
\label{f:logtr}
\EEQ

As for the standard Parisi (``one-group'') Ansatz, the
self-consistency
conditions for the large $N$ saddle point read
\BEQ
Q^{\sigma\sigma'}_{ab} =\left\langle s^{i_\sigma}_a
s^{i_{\sigma'}}_b
\right\rangle
\EEQ
where $\sigma, \sigma' \in \{+,-\}$, and the average is taken over
the
Boltzmann factor in Eq.~(\ref{f:logtr}).
Here an index $i_+$ is restricted to $[1,m+K]$, while
$i_-\in [m+K+1,m]$.
Solving for $Q_{ab}$, $A_{ab}$ and $C_{ab}$ (cf.
Eqs.~(\ref{Qpm},\ref{Qpp}))
we may also write
\begin{eqnarray}
\label{selfcons2G_1}
Q_{ab} &=& \frac{1}{4}\left\langle \left(s^{i_+}_a + s^{i_-}_a\right)
 \left(s^{i_+}_b + s^{i_-}_b\right) \right\rangle,\\
A_{ab} &=&
\frac{K}{2} ~
\left\langle s^{i_+}_a\left(s^{i_+}_b-s^{i_-}_b\right)  +
 \left(s^{i_+}_a-s^{i_-}_a\right) s^{i_-}_b \right\rangle \\
C_{ab} &=& K^2
\left\langle (s^{i_+}_a -s^{i_-}_a)(s^{i_+}_b -s^{i_-}_b) \right\rangle,
\label{selfcons2G_1C}
\end{eqnarray}
which resembles Eqs.~(\ref{QOP}-\ref{COP}). The detailed connection between the two sets of equations is established in App.~
\ref{app:phi_m}.

The log-trace term $\phi_m$ in Eq.
(\ref{f:logtr})
can be re-expressed  in two equivalent ways which will be helpful
to make the connection
with the generalized cavity approach and the counting of TAP
states, respectively.
 Technical details of the derivation can be found in App.~\ref{app:phi_m}.


Following Parisi and Potters \cite{PP}
one  obtains the expression
\begin{eqnarray}
\label{f:phim_pp}
&&e^{n\beta\phi_m}=
2^{nm}
{\int_{-\mbox{\scriptsize i}\infty}^{\mbox{\scriptsize i}\infty}}
\prod_{a}\frac{dx_a}{2\pi{\rm
i}}\int_{-1}^{1}\prod_{a}\frac{dm_a}{(1-m_a^2)}
\\
\nn
&&\times\exp\left\{-\sum_{a} \left[x_a\tanh^{-1}m_a
+\frac{m}{2}\log(1-m_a^2)\right]\right.
\\
\nn
&&\hspace*{.5cm}\left.+\beta^2 \sum_{ab}\frac{1}{2}
x_aQ_{ab} x_b+\frac{1}{2}m_aC_{ab} m_b+m_aA_{ab} x_b\right\},
\end{eqnarray}
where the sum over $(a,b)$ also includes diagonal terms,
$Q_{aa}\equiv Q$, $A_{aa}\equiv A$, $C_{aa}\equiv C$.
We will see  in Sec.~\ref{s:TAPcount} that this form also emerges
 from a direct
counting of TAP solutions.

In this representation, the self-consistency equations (\ref{selfcons2G_1})
 can be cast into the form
\BEA
\label{f:Qab_sc}
&&Q_{ab} = \left<m_a m_b\right>,
\\
\label{f:Aab_sc}
&&A_{ab} +\delta_{ab}(1-Q_{aa})= \left<m_a(x_b-m m_b)\right>,
\\
\label{f:Cab_sc}
&&C_{ab} -\delta_{ab}[ 2A_{aa}+m(1-Q_{aa})]
\\
&&\hspace*{2cm}= \left<(x_a-m m_a)(x_b - m m_b)\right>,
\nn
\EEA
where the average $\left<\ldots \right>$ is taken over the measure
in
Eq.~(\ref{f:phim_pp}).

In appendix \ref{app:phi_m} we show that the field $x_a$ is in a certain sense a twofold Hubbard
Stratonovich transformation of the spin variables $S_a=\sum_i
s_a^i$. Its average magnetization is therefore $m m_a$, and
the terms $x_a-m m_a$ can be thought of as magnetization fluctuations in replica
space. Indeed this furnishes an intuitive understanding of the off-diagonal part ($a \neq b$) of  Eqs.~(\ref{f:Qab_sc}-\ref{f:Cab_sc}) and supports our interpretation of the overlap matrices.
The extra diagonal terms ($a=b$) on the left hand side of Eqs.~(\ref{f:Aab_sc}-\ref{f:Cab_sc}) arise due to the fragility of the marginal states, as we will see
in Sec. \ref{s:cav} from an alternative derivation.

In appendix \ref{app:phi_m} we derive the equivalent expression
\begin{eqnarray}
\label{f:phim_cav}
&&e^{n\beta\phi_m}=\frac{1}{\sqrt{\det{\cal M}}}
\int \prod_{a=1}^n\frac{ dh_a dz_a}{2\pi}
\\
\nn
&&\exp\Bigl\{
\sum_a \left[m \log 2\cosh(\beta h_a) +
\tanh(\beta h_a) \beta z_a\right]
\\
\nn
&&\hspace*{3.8 cm}
-\frac{1}{2}\sum_{ab}{\bf {\underline{\xi}}}_a^\dag \cdot {\cal M}^{-1}_{ab}
\cdot
{\underline{\xi}}_a
\Bigr\},
\end{eqnarray}
where ${\underline \xi}_a^\dag \equiv (h_a,z_a)$ and
${\cal M}$ is the $2n\times 2n$ covariance matrix
\begin{equation}
{\mathbf{\cal M}}=\Bigl(\overbrace{
\begin{array}{c}
 \mathbf{Q} \\
 \mathbf{A}\end{array}
}^n
 \overbrace{
\\
\begin{array}{c}
\mathbf{A} \\
\mathbf{C}
\end{array}
}^n
\Bigr).
\end{equation}
Moreover, introducing magnetization and soft mode variables (cf.~Eqs.~(\ref{app:vartrafo1},\ref{app:vartrafo2})),
\BEA
\label{vartrafo1}
\tilde{m}_a&=&\tanh(\beta h_a),\\
\label{vartrafo2}
\delta m_a&=&\frac{\beta z_a}{\cosh^2(\beta h_a)},
\EEA
the self-consistency equations
(\ref{selfcons2G_1}-\ref{selfcons2G_1C}) take the simple form
\begin{eqnarray}
\label{SCsimpleQ}
Q_{ab} &=& \left\langle \tilde{m}_a\tilde{m}_b \right\rangle,\\
A_{ab} &=&\left\langle \tilde{m}_a\delta m_b \right\rangle,\\
C_{ab} &=& \left\langle \delta m_a\delta m_b \right\rangle,
\label{SCsimpleC}
\end{eqnarray}
where averages are over the measure defined by the integrand in Eq.~(\ref{f:phim_cav}). We will use this form in the discussion of the quenched computation in Sec.~\ref{s:quenched}.

\subsection{Annealed approximation}
\label{ss:2G_ann}
In the following we will focus on the annealed
approximation, which corresponds to averaging
the two group partition function ${\overline {Z_J^m}}$ instead of
its logarithm.
This is known to be an exact description at high enough free energy densities, $f>f^\star$,\cite{BMJPC81}
cf. Eq.~(\ref{f:eq_ann}) below. Technically, this approximation corresponds to reducing
the overlap matrix to its diagonal part ($a=b$) described by $\{Q,A,C\}$, and setting all off-diagonal elements to zero.

The formalism for a quenched computation with continuous RSB is reviewed and physically interpreted in Sec. \ref{s:quenched}.

Integrating out $z_a$ in the annealed version of Eq.~(\ref{f:phim_cav}), one finds
\BEA
\label{f:mu(h)}
e^{\beta \phi_m}&=&
\int_{-\infty}^\infty\frac{dh}{\sqrt{2 \pi Q}}e^{-\frac{h^2}{2Q}}
(2\cosh\beta h)^m
\\
&&\hspace*{.2cm}\exp\Bigl [\frac{A}{Q}\beta h \tanh\beta h
+\frac{\beta^2}{2}\frac{QC-A^2}{Q}\tanh^{2}\beta h
\Bigr ].\nn
\EEA
The same result is
obtained by integrating out $x_a$ in the annealed version of Eq.~(\ref{f:phim_pp})
and using the relation
$m_a =\tanh \beta h_a$  between magnetization $m_a$ and local field $h_a$.
The integrand on the right hand side has the interpretation of a probability distribution
of local fields (up to a normalization):

\BEA
\label{Pofhann}
P_{\rm ann}(h)&=&
\frac{1}{\cal N}
\exp\Bigl\{m\log\left(2\cosh\beta h\right)-\frac{h^2}{2Q}
\\
\nn
&&\hspace*{.5cm}+\frac{A}{Q}\beta h \tanh\beta h
+\frac{\beta^2}{2}\frac{QC-A^2}{Q}\tanh^{2}\beta h
\Bigr\}
\EEA
The joint distribution of magnetizations $\tilde{m}$ and soft mode components $\delta m$ can be similarly obtained from normalizing the replica-diagonal version of the measure in Eq.~(\ref{f:phim_cav}) and changing variables according to Eqs.~(\ref{vartrafo1},\ref{vartrafo2}).

In the annealed approximation Eqs. (\ref{f:Qab_sc}-\ref{f:Cab_sc}) take the form  (performing the Gaussian averages over $x$)
\BEA
&&
Q=\left<\tanh^2\left(\beta h\right)\right>,
\label{saddleQz}
\\
\label{saddleA2z}
&&
A+1-Q=-A-m Q
+\frac{\left<h \tanh \left(\beta h\right)
\right>}{\beta Q},
\\
&&
\label{saddleCz}
C-2A-m(1-Q)=m^2Q+2mA+\frac{A^2}{Q}
\\
&&\hspace*{.5 cm}
-2\left(m+\frac{A}{Q}\right)
\frac{\left<h\tanh\left(\beta h\right)\right>}{\beta Q}
-\frac{1}{\beta^2 Q}\left(1-\frac{\left<h^2\right>}{Q}\right),
\nn
\EEA
the average $\left<\ldots\right>$ now denoting an integral over
$P_{\rm ann}(h)$.  Eqs. (\ref{Pofhann}-\ref{saddleCz}) allow one to
find the annealed solution $\{Q,A,C\}$ easily, e.g., by iteration.

The above equations turn out to admit two solutions,\cite{CLPR03} only
 one of which (with $A,C\neq 0$) is physical, as discussed further in
 Secs.~\ref{s:TAPcount} and \ref{s:stab}.

Bray and Moore showed in Ref.~[\onlinecite{BMJPC81}] that this annealed
solution is stable with respect to continuous replica symmetry
breaking (onset of correlations between typical metastable states) as
long as $f>f^\star=f(m^\star)$ where $m^\star$ satisfies the condition
\BEA
1&=& \beta(1-Q) + \beta\left(A+\frac{m^\star Q}{2}\right)\nn\\
&&+\beta\sqrt{Q\left(C+m^\star A+\frac{m^{\star 2}}{4}Q-2A\right)}.
\label{f:eq_ann}
\EEA

\subsection{Complexity and direction of the marginal mode}

\begin{figure}[b]
\resizebox{0.49\textwidth}{!}{\includegraphics{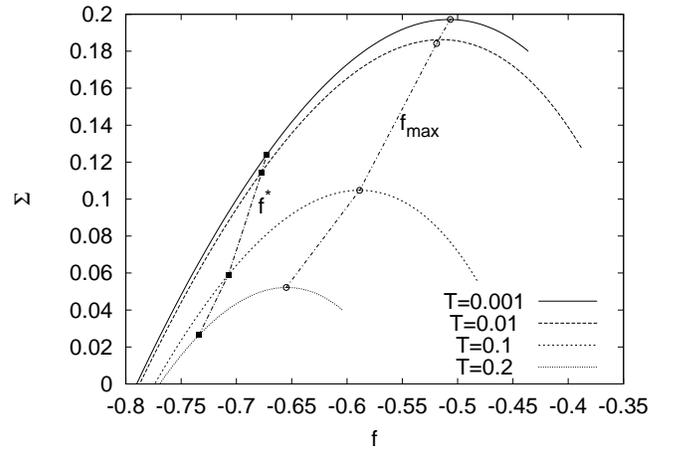}}
\caption{Complexity curves in the annealed approximation. Only the
data for $f>f^\star$ is exact. $f_{\rm max}$ corresponds to the states
of maximal complexity at a given temperature.  The high energy part of
the curves is not fully shown (they extend up to $f_{\rm hbe}$ where
$\Sigma(f_{\rm hbe}=0$).)  }
\label{fig:complexity}
\end{figure}

Evaluating the replicated free energy $\Phi_{2G}$ [Eq.~(\ref{Eq_Phi})]
within the annealed approximation, one eventually obtains the complexity as
its Legendre transform:
\BEQ
\Sigma(f) =- \beta m \Phi_{\rm 2G}(m)
+m\beta f.
\label{SLT}
\EEQ
 In the annealed approximation the above observables read
\BEA
f(m)& =& \frac{\p m \Phi_{\rm 2G}(m)}{\p m}=-\frac{\left<\log 2 \cosh\beta h\right>}{\beta}
\\
&&
-\frac{\beta}{4}\Bigl[\left(1-Q\right)^2-2 m Q^2
-4AQ\Bigr],\nn
\\
\Sigma(m) &=& \beta m^2 \frac{\p \Phi_{\rm 2G }}{\p m} =\beta\phi_m
-m
\left<\log 2 \cosh\beta h
\right>
\label{f:sigma_ann}
\\
&&
\nn
+\frac{\beta^2}{4}\left[ m^2Q^2
-4A(1-Q) -2 (A^2 + Q C) \right],
\EEA
that are used to compute the complexity curves of
Fig.~\ref{fig:complexity}.

\begin{figure}[t]
\resizebox{0.49\textwidth}{!}{\includegraphics{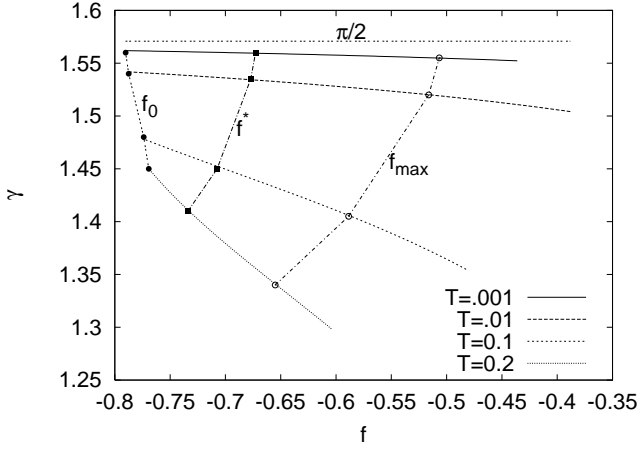}}
\caption{Angle between soft mode and magnetization of typical marginal
states. The parameter $f_0$ is the value of the free energy density at
which the complexity of marginal states computed in the annealed
approximation goes to zero ($f_0$ is not equal to the true equilibrium
value $f_{\rm eq}$ since the approximation breaks down below
$f^\star$).  $f_{\rm max}$ denotes the free energy density with
maximal complexity (see Fig.  \ref{fig:complexity}).}
\label{fig:gamma}
\end{figure}

In Fig.~\ref{fig:gamma} we plot the angle $\gamma$ between the soft
mode and the magnetization vector in configuration space, cf.~
Eq.~(\ref{f:gamma}), versus free energy density for various
temperatures. Only the part of the curves to the right of $f^\star(T)$
is exact, whereas the low energy part is an annealed approximation to
the regime where states are correlated among each other.
Nevertheless, there is a clear tendency to increasing towards
orthogonality, $\gamma\rightarrow \pi/2$, both as $f$ is approaching
the lower band edge of the $f$ interval, and as $T\rightarrow 0$. We
will derive this behavior explicitly in the low $T$ limit in
Sec.~\ref{s:lowT}.

\section{Counting TAP states:  breaking of BRST symmetry
and equivalence to the two group Ansatz}
\label{s:TAPcount}
In this section we briefly review the counting of TAP states,
defined as stationary points (minima) of the TAP free energy
\begin{eqnarray}
F_{\rm TAP}(\{m_i\})&=&-\frac{1}{2}\sum_{ij}J_{ij}m_im_j+\frac{1}{\beta}\sum_i
f_0(q,m_i),\nn\\
\label{f:FTAP}
\end{eqnarray}
with
\begin{eqnarray}
f_0(q,m_i)&=&\frac{1}{2}\ln(1-m_i^2)+m_i\tanh^{-1}m_i
\\
\nn
&&-\ln 2-\frac{\beta^2}{4}(1-q)^2.
\end{eqnarray}

In order to select states at a given free energy density we
weight the TAP states (labeled by $\alpha$) with an exponential factor:
\begin{eqnarray}
{\cal Z}_J(\beta;m)&\equiv&\sum_{\alpha}e^{-m\beta F_{\rm TAP}
(\{m_\alpha)\}}.
\label{f:TAP_partition}
\end{eqnarray}
We can rewrite this sum in the integral
representation~\cite{Annibale}
\begin{eqnarray}
\label{action}
{\cal Z}_J(\beta;m)&=& \int {\cal D}m {\cal D}x {\cal
D}\overline{\psi}{\cal D}\psi
\\
\nn
&&\exp\Bigl\{\beta \Bigl[-\sum_i x_i
\partial_{m_i} F_{\rm TAP}(\{m_i\})
\\
\nn
&&\hspace*{1cm}
+\sum_{ij}\overline{\psi}_i\psi_j\partial_{m_i}\partial_{m_j} F_{\rm
TAP}(\{m_i\})
\\
\nn
&&\hspace*{1cm}
- m F_{\rm TAP}(\{m_i\})\Bigr]\Bigr\}.
\end{eqnarray}
One then proceeds by replicating the above expression, averaging
over
the Gaussian bond disorder, decoupling quartic terms in
$x,m,\psi,\overline{\psi}$ by Lagrange
multipliers. Integrating out the fermionic fields
and imposing saddle point conditions, one obtains the
expression~\cite{fn4}

\BEA
\label{Cavagna}
&&\overline{{\cal Z}^n(\beta;m)} = \prod_a \int_{-1}^1
\frac{dm_a}{1-m_a^2} \prod_a \int_{-i\infty}^{i\infty}
\frac{dx_a}{2\pi i}
\\
\nn
&&\hspace*{.4cm }
\exp \Bigl\{
N\Bigl[
-\sum_a x_a \tanh^{-1}(m_a) - m\sum_a f_0(Q_{aa},m_a)
\nonumber\\
&&\hspace*{1.4cm }
-\frac{\beta^2}{2} \sum_{ab} G_{ab}
-\frac{\beta^2}{2} \sum_a 2A_{aa}(1-Q_{aa})
\nonumber\\
&&\hspace*{1.4cm }
+\frac{\beta^2}{2}\sum_{ab}
\left(x_aQ_{ab}x_b+x_a
(2 A_{ab}+mQ_{ab})m_b
\right.
\nn
\\
\nn
&&\hspace*{2.4cm}\left.+m_a(C_{ab}+2mA_{ab}+m^2Q_{ab})m_b\right)
\Bigr]\Bigr\}.  \EEA
with
\BEA
\nn
G_{ab}\equiv
2(A_{ab}^ 2+ Q_{ab}~C_{ab})+4mA_{ab}~Q_{ab}
+m^2Q_{ab}^2
\\
\label{G}
\EEA

This should be optimized with respect to $Q_{ab}, A_{ab},
C_{ab}$,
leading to the saddle point equations
\begin{eqnarray}
\label{SPTAP1}
Q_{ab}&=&\langle m_a m_b \rangle,\\
\label{SPTAP2}
A_{ab}+\delta_{ab}(1-Q_{aa})&=&\langle x_a m_b \rangle,\\
\label{SPTAP3}
C_{ab}-\delta_{ab}(2A_{aa}+m(1-Q_{aa}))&=&\langle x_a x_b \rangle.
\end{eqnarray}
As in the previous section, these equations admit two solutions,
i.e., there are
two possible saddle points for the integrand in Eq.~(\ref{action}):
(i) one with $A=C=0$, conserving the BRST symmetry of the action (\ref{action}), and
(ii) a saddle point with $A\neq 0$, $C\neq 0$ which spontaneously breaks this
symmetry. We will discuss in Sec.~\ref{s:stab} that only the second solution is physical.

\subsection{Equivalence with the 2G Ansatz}
Taking the logarithm of Eq.~(\ref{Cavagna}) we obtain
\BEQ
\label{f:Phi_tap}
{\cal F}(\beta;m)\equiv {\overline {\log
{\cal Z}_J}}=\lim_{n \to 0}\frac{1}{n}\log
{\overline {{\cal Z}_J^n}(\beta;m)}
\EEQ
where $m$ is the factor in the weight of Eq.
(\ref{f:TAP_partition}).
It takes the same role as the total
number of replicas in the two groups in the previous section.
Again, a Legendre transform with respect to $m$ yields the complexity function.

Comparing the thermodynamic potentials of~Eq.  (\ref{f:Phi_tap}) and
Eq. (\ref{Phi2G}) (evaluated according to Eq.~(\ref{f:phim_pp})) we
establish their complete equivalence after shifting the integration
variable to $\tilde{x}_a = x_a-mm_a$ in Eq.~(\ref{f:phim_pp}). As
explained in section~\ref{s:2G} this shift amounts to using natural
variables for magnetization fluctuations (see also appendix
\ref{app:phi_m}).

From the present derivation we obtain a better understanding
of the auxiliary fields $x_a$ in the saddle point
equations (\ref{SPTAP1}-\ref{SPTAP3}), which are obviously
equivalent to Eqs. (\ref{f:Qab_sc}-\ref{f:Cab_sc}) obtained from the two group Ansatz.
As one may expect from
the functional integral (\ref{action}) the variables $x_a$ are prone to fluctuate
most strongly along the softest mode of the TAP solutions, and this leads to a rather natural
 interpretation of the saddle point equations (\ref{SPTAP1}-\ref{SPTAP3})
 for $a\neq b$:
$A_{ab}$ and $C_{ab}$ describe
correlations
between the soft mode directions and the local magnetizations, as also suggested by the two group Ansatz.

For $a=b$ extra terms arise. This is because $x_a$ does not exactly describe
 soft mode directions in the same sense as $m_a$ describes magnetizations.
While this is immaterial for inter-state correlations, it introduces correction
terms when intra-state correlations are considered.
In the next subsection, this will become clearer from a direct derivation of these contributions from generalized Ward identities.


\subsection{BRST symmetry breaking and the generalization of
 Ward identities.}
The action in the exponent of the integrand in Eq.
(\ref{action}) is invariant under the fermionic
BRST symmetry
\cite{BRST}
\BEA
\delta m_i = \epsilon \psi_i,\ \ \delta x_i = -\epsilon \frac{m}{2}
\psi_i,
  \ \delta {\overline \psi_i} = -\epsilon x_i,  \ \ \delta \psi_i
= 0
\label{f:brst}
\EEA
If the dominant states are stable minima (cf. Sec.
\ref{ss:genmin}),
the BRST-symmetry of the
action (\ref{action}) is not broken by the saddle point Eq.
(\ref{Cavagna}),  and Ward
identities
\BEA
\langle m_i x_i
\rangle &=& \langle \psi_i \overline{{\psi}_i}\rangle,
\label{f:w1}
\\
\label{f:w2}
 \langle x_i x_i \rangle &=&
- m~\left<\psi_i{\overline \psi_i}\right>,
\EEA
hold. On the macroscopic level (upon average over sites $i$), these impose that the order
parameters $A_{ab}=C_{ab}=0$.
This can be appreciated by noting that Eqs.~(\ref{f:w1},\ref{f:w2}) reproduce the diagonal
saddle point equations  (\ref{SPTAP2},\ref{SPTAP3}), considering
 that $N^{-1}\sum_i \left<\psi_i{\overline \psi_i}\right>=1-Q$.

 If instead the majority of states
are marginally stable states (i.e., minima merging with saddles of
rank
one) the BRST symmetry is broken, translating into  $A_{ab}\neq 0$, $C_{ab}\neq0$.
The Eqs.~(\ref{SPTAP1}-\ref{SPTAP3})
result from a (macroscopic) generalization of  the above Ward identities.
One can derive them from direct
inspection of the sum over TAP states with a regularization.
As we did for the toy model of Sec.
\ref{s:toy} [Eq.~(\ref{constraint})]
we add a regularizing term $\exp(\lambda_X X(\{m_j\}))$ to the weight in Eq.~(\ref{f:TAP_partition}). As explained in App.~\ref{app:regularization}, this procedure selects minimally stable states with a susceptibility diverging as
$1/\lambda_X$ as $\lambda_X \rightarrow 0$:
\BEQ
{\cal Z}_J(\beta;m,\lambda_X)=\sum_{\alpha=1}^{{\cal N}_{\rm sol}}
e^{-m\beta F_{TAP}(\{m_\alpha\})
+\lambda_X X(\{m_\alpha\})}.
\label{f:regul}
\EEQ

The quantity  $X$ is an arbitrary, extensive, symmetric function of the
average site magnetizations subject only to the constraint that its gradient be
non-orthogonal to the soft mode of typical marginal states:
\BEQ
{\vec \nabla}_{m} X({\vec m})\cdot {\vec{\delta m}}\neq 0.
\label{conditionSK}
\EEQ
This regularization explicitly lifts the marginality
throughout the computation. Sending the control
parameter
$\lambda_X$  eventually to zero allows us to recover the marginal states.

When the TAP equations are perturbed by external fields there is no
guarantee that
the number of solutions
at a given value of the free energy density is conserved.
Indeed only the set of TAP solutions counted by
the BRST symmetric saddle point is robust to perturbations:\cite{CLPR03}
the number of
solutions at any free energy density is conserved and there is a one-to-one
correspondence between states before and after the perturbation.
In the case of broken BRST symmetry the correspondence is lost and
solutions
can appear and disappear at any free energy density. This
fragility
is related to the marginality of the states.
The regularization Eq.~(\ref{f:regul}) fixes this problem by lifting
the marginality. As a consequence the number of selected states is conserved under perturbations (for fixed parameters $f$
{\em and} $\lambda_X$), but at the same time the BRST symmetry is explicitly broken.

We will now derive the generalizations of the Ward identities
(\ref{f:w1},\ref{f:w2}). Moreover, by showing their equivalence with
the self-consistency Eqs.  (\ref{SPTAP1}-\ref{SPTAP3}), we shall
confirm the interpretation of the order parameters of the two group
Ansatz put forward in Sec.~\ref{ss:2gop}.

We consider sums over TAP states which are dominated by shallow minima
due to regularization. Further, we regard the states as functions of
small perturbing fields $\{h_k\}$, in the sense that
$m_\alpha(\{h_k\})$ denotes the unique solution of $\partial_{m_i}
F_{\rm TAP}(\{m_j\}) =h_i$ in the vicinity of the unperturbed state
$m_\alpha(\{h_k=0\})$.

Since the free energy Hessian possesses a soft mode along the
direction $\delta m_i \propto \zeta_i$ (normalized by
$N^{-1}\sum_i\zeta_i^2=1$), the latter dominates the response to a
perturbing external field.  The regularization selects states whose
susceptibility along the soft mode is finite, $\chi_{\rm
soft}=\sum_{ij}\zeta_i\chi_{ij}\zeta_j =(g \lambda_X)^{-1}$, where $g$
is a selfaveraging constant, as shown in
App.~\ref{app:regularization}.

The coupling of perturbing external fields $\{h_i\}$ to the soft mode is given
 by their projection on the soft mode, $\sum_i\zeta_i h_i$.
In particular, restricting to the response along the soft mode, we find
\BEA
\frac{\partial m_i}{\p h_k}\approx \frac{\zeta_i\zeta_k}{g \lambda_X}
\EEA
and thus, in the limit of vanishing regularizer one finds a finite
limit for the derivative
\BEA
\frac{1}{\beta}\lim_{\lambda_X\to 0}\frac{\partial\lambda_X X }{\partial
h_k}=\frac{1}{\beta g}
\left(\sum_i \frac{\partial X}{\partial m_i}\zeta_i \right) \zeta_k
\equiv  \delta m_k
\label{f:soft_mode}
\EEA
Note that the so defined vector $\{\delta m_k\}$ is proportional to
the soft mode $\{\zeta_k\}$ and independent of $X$ (since the constant
$g$ is proportional to $X$, and only the gradient of $X$ in the
direction of $\vec{\zeta}$ matters). This same vector will appear
again in the cavity method (see next section~\ref{s:cav}). It gives a
precise meaning to the amplitude of the soft mode appearing in the
heuristic derivation of the two group Ansatz, cf.,
Eq.~(\ref{2G:pheno}).

With the above observation, we are able to derive the following
generalized Ward identities (see Appendix~\ref{app:ward} for the
details):
\BEA
\label{f:Ward1}
\langle m_i^a x_i^b
\rangle &=& \delta_{ab} \langle \psi_i^a
{\overline{\psi}}_i^b\rangle +
\langle m_i^a \delta m_i^b \rangle,
\\
 \langle x_i^a x_i^b \rangle &=&
- m~\delta_{ab}\left<\psi_i^a{\overline \psi}_i^b\right>
+ \langle \delta m_i^a \delta m_i^b \rangle\\
&&\hspace*{3cm} -2  \delta_{ab}\left
\langle m_i^a\delta m_i^b
\right\rangle.\nn
\label{f:Ward_gen2}
\EEA
These equations are indeed equivalent to Eqs.
(\ref{SPTAP2}-\ref{SPTAP3})
provided that we make the identification
$A_{ab}=\left\langle
m_a\delta m_b\right\rangle$, $C_{ab}=\left\langle \delta m_a\delta
m_b
\right\rangle$, as suggested by our interpretation of $A$ and $C$ as describing correlations among soft mode directions and magnetizations.

In the absence of soft modes
the limit $\lambda_X\rightarrow 0$ cancels the terms containing
${\delta\vec{ m}}$ and the BRST-Ward identities (Eqs.
(\ref{f:w1},\ref{f:w2}))
are recovered automatically.

\section{Generalized cavity approach and the two group Ansatz}
\label{s:cav}
\subsection{Cavity with marginal states revisited}
In the standard cavity approach one writes recursion relations for
the local (cavity) field $h_0$ acting on an added site $0$ in terms of the local fields $h_i$ acting on its
$k$ neighboring sites when site $0$ is absent. Here, $k+1$ is the connectivity of the lattice.
In
Ising systems, the cavity field is a sum of {\em messages}
 $u_i$,\cite{MPEPJB01}
\begin{eqnarray}
\label{recursions}
h_0 &=& \sum_{i=1}^k u_i(h_i;J_{0i}),\\
u_i(h_i;J_{0i})&=& \beta^{-1}\tanh^{-1}
[\tanh(\beta h_i) \tanh(\beta J_{0i})],
\end{eqnarray}
where
$J_{0i}$ are the quenched bonds coupling the cavity spin $0$ to its neighbors.
The free energy gain for the addition of a spin at site 0 is
\begin{equation}
\label{deltaF}
\exp(-\beta \Delta F)=2\cosh(\beta h_0)\prod_{i=1}^k
\frac{\cosh(\beta
J_{0i})}{\cosh(\beta u_i)}.
\end{equation}

 Around the considered free energy density $f$, the density of metastable states
 is assumed to grow exponentially as $\rho(F)\propto \exp[m(f)(F-F_0)]$
where $m(f)=\p \Sigma(f)/\p f$ is the local slope of the complexity function
(in other words, $m$ is the Legendre conjugate of $f$), $F=fN$ and $F_0$ is an
arbitrary reference value (to be absorbed into the normalization constant).
 In order to
determine the average shift $\Phi_{\rm cav}(m)$ of this distribution when a site is added,
one averages the 'reweighting' factor $\exp(-\beta m \Delta F)$ over cavity iterations  to obtain~\cite{MPEPJB01}
$\exp[-\beta m\Phi_{\rm cav}(m)]\equiv \left<\exp[-\beta m \Delta F]\right>_{\{h_i, J_{0i}\}}$.

In the case of the SK model the connectivity goes to infinity
$N\equiv k+1\rightarrow \infty$ and the single bond strength tends
to zero as $\overline{J^2_{0i}} = 1/N$.
The above relations thus simplify. In
particular, we have $u_i\approx J_{0i}m_i$ with $m_i \equiv
\tanh(\beta h_i)$. The free energy shift due to a spin addition is
\BEA
\label{deltaFSK}
\exp(-\beta \Delta F)&=&2\cosh(\beta h_0)
\exp\left[\frac{\beta^2}{2}\sum_{i=1}^N J_{0i}^2(1-m_i^2)\right]\nn\\
&=&2\cosh(\beta h_0)
\exp\left[\frac{\beta^2}{2}\langle 1-m_i^2\rangle \right],
\EEA
where $\langle.\rangle$ denotes a site average, and we have used the fact that in the large $N$
 limit the second term does not fluctuate.

If a marginal mode is present, this standard method fails since the
addition of a spin has an anomalously strong impact on the system,
rendering previous states unstable.  In order to circumvent this
difficulty, we regularize the problem once again by reweighting the
states with a factor $\exp[\lambda_X X(\{m_i\})]$, in addition to the
standard reweighting $\exp[-\beta m F(\{m_i\})]$ with respect to free
energy which selects a certain free energy density. Eventually, we
will take $\lambda_X\rightarrow 0$ to recover the marginal states. As
in earlier sections, the extensive observable $X$ is an arbitrary
symmetric function of the magnetizations, subject to the requirement
(\ref{conditionSK}).  This method was originally introduced by
Rizzo.~\cite{RizzoJPA05} Here we go beyond his analysis by including
the selection of a given free energy density, and establishing the
precise connection with the two group Ansatz. Further, we provide a
clear interpretation of the auxiliary cavity fields that need to be
introduced.

We expect from the toy models of Sec. \ref{s:toy}
that the regularization scheme will select minimally stable states with the lowest eigenvalue of their free energy Hessian being
proportional to $\lambda_X$ (for a derivation in the present case see App.~\ref{app:regularization}).
 The change of $X$ upon addition of a site
is most conveniently computed as a derivative of the free energy
change with respect to the field $h_X$ conjugated to the observable $X$,
\begin{equation}
\Delta X= -\frac{d \Delta F}{d h_X}.
\end{equation}
The regularization is implemented in the cavity approach by reweighting each cavity iteration by $\exp[\lambda_X \Delta X]$.
 Note that the exponent remains finite in the limit $\lambda_X
\rightarrow 0$, because the susceptibility of the soft mode diverges as $1/\lambda_X$.
More precisely we have,
\begin{equation}
\label{Xshift}
\lim_{\lambda_X\to  0}\lambda_X \Delta X
=
-\sum_{i=0}^N \frac{d \Delta F}{d h_i} \lambda_X
\frac{dh_i}{dh_X}= -\sum_{i=0}^N \frac{d \Delta F}{d h_i}
\beta z_i,
\end{equation}
where
\BEQ
\label{def:zcav}
z_i \equiv \beta^{-1} \lim_{\lambda_X\to 0}\lambda_X
\frac{dh_i}{dh_X}
\EEQ
 are finite fields proportional to the  component $\Delta h_i$ of the local field fluctuation arising from a soft mode excitation. The fields $z_i$ are in fact independent of the choice of $X$, as shown in App.~\ref{app:regularization}.

Deriving the recursion relation~(\ref{recursions}) with respect to $h_X$ we obtain a relation for $z_0$
\begin{eqnarray}
\label{recursionz}
z_0 =\frac{dh_0}{dh_X}&=& \sum_{i=1}^N J_{0i} \beta(1-m_i^2)z_i
 \\
\nn
&\equiv& \sum_{i=1}^N
J_{0i} ~\delta m_i,
\end{eqnarray}
where we introduced the soft mode in the magnetizations
\BEQ
\delta m_i \equiv \frac{dm_i}{dh_i} z_i =
\beta(1-m_i^2)z_i.
\EEQ
Note the correspondence with the definition (\ref{f:soft_mode}), which is most transparent if we notice that
\BEQ
\delta m_i =
\beta^{-1}\lim_{\lambda_X\to 0} \lambda_X \frac{dm_i}{dh_X},
\EEQ
using the definition (\ref{def:zcav}).

For the SK model, the shift of the regularizer, Eq.~(\ref{Xshift}), is readily evaluated
using Eq.~(\ref{deltaFSK})
\begin{eqnarray}
\label{XshiftSK}
\lim_{\lambda_X\to 0}
\lambda_X \Delta X &=& \tanh(\beta h_0) \beta z_0
-\sum_{i=1}^N \tanh(\beta u_i) \frac{du_i}{dh_i} \beta z_i \nn
\\
&=&
\beta m_0 z_0-\beta^2 \sum_{i=1}^N J_{0i}^2 m_i \beta(1-m_i^2)
z_i\nonumber\\
&=&\beta m_0 z_0-\beta ^2 \langle m_i \delta m_i\rangle,
\end{eqnarray}
where  $m_0=\tanh \beta h_0$. Similarly as in Eq.~(\ref{deltaFSK}), we may neglect the fluctuations of the second term in the large $N$ limit.

\subsection{Free energy and self-consistency equations}

Putting elements together, we obtain the regularized free energy shift $\exp[-\beta m \Phi_{\rm cav}]$
upon a site addition
by averaging the two
reweighting factors $\exp[-m\beta \Delta F]\exp[\lambda_X\Delta X]$ over all possible random configurations $\{h_i,z_i\}$ of the neighboring
cavity fields, as well as over the random couplings $J_{0i}$ 
\BEA
\label{phiofm}
&&\exp[-\beta m \Phi_{\rm cav}]
\\
&&\hspace*{.5cm}
= \langle \exp (-\beta m \Delta F)
\exp(\lambda_X \Delta X) \rangle_{\{h_i,z_i,J_{0i}\}}.
\nn
\EEA

The first term has the standard interpretation of a shift of the exponential distribution
of states. The second term equals 1 if the considered states are stable.
In the marginal case, it captures information about the probability for site
additions to render existing states unstable or to make new marginal states emerge.

We may use the central limit theorem to infer from Eqs.~(\ref{recursions},\ref{recursionz}) that $h_0(\{h_i,z_i,J_{0i}\})$ and $z_0(\{h_i,z_i,J_{0i}\})$ are Gaussian variables with covariance matrix
\begin{equation}
{\cal M}=\left(\begin{array}{cc}
 Q^\star & A^\star\\  A^\star & C^\star\end{array}\right),
\end{equation}
where
\begin{eqnarray}
Q^\star&=&{\overline{\left(\sum_{i=1}^N J_{0i}m_i\right)^2}}
=
 \langle m_i^2 \rangle,
\\
A^\star&=& {\overline{\sum_{i=1}^N J_{0i}m_i \sum_{i=1}^N
J_{0i}\delta m_i}}
\\
 \nn
&&
\hspace*{1cm}= 
\langle m_i \delta m_i \rangle = 
\langle m_i \beta(1-m_i^2)  z_i\rangle,
\\
C^\star&=&
{\overline{\left(\sum_{i=1}^N J_{0i}\delta m_i\right)^2}}
\\
\nn &&\hspace*{1cm}=
\langle (\delta m_i)^2\rangle  = 
\langle [(1-m_i^2) \beta z_i]^2\rangle,
\end{eqnarray}
and $\langle.\rangle$ denote site averages.
We can then reexpress Eq.~(\ref{phiofm}) as
\BEA
\label{phiofm1}
&&\exp[-\beta m   \Phi_{\rm cav}]\\
&&\hspace*{.5cm}= \exp\left[m\frac{\beta^2}{2}(1-Q^\star)-\beta^2 A^\star\right]
\\
&&\int \frac{dh_0 dz_0}{2\pi [\det{\cal M}]^{1/2}}
\exp\left[-\frac{1}{2}~~ {\underline{\xi}}_0^\dag\cdot {\cal M}^{-1}
\cdot {\underline{\xi}}_0\right]
\nonumber
\\
\nn
&&\hspace*{2.5cm} [2\cosh(\beta h_0)] ^m \exp\left[\tanh(\beta h_0)
\beta z_0\right]
\end{eqnarray}
with ${\underline{ \xi}}^\dag_0 \equiv (h_0,z_0)$.
The reweighting terms can alternatively be seen as describing the relative probability of a cavity
 configuration $\{h_i,z_i,J_{0i}\}$ to occur, given a fixed free energy after the site addition. With this interpretation in mind,
the probability distribution to find local fields $h_0$ and soft mode components
$z_0$ on site $0$ can be read off from Eq.~(\ref{phiofm1}) as
\begin{eqnarray}
P(h_0,z_0) &=& {\cal N}^{-1}
\exp\left[-\frac{1}{2} ~{\underline{ \xi}}^\dag_0 \cdot {\cal M}^{-1}
\cdot {\underline{ \xi}}_0\right]
\\
&&(2\cosh(\beta h_0)) ^m \exp\left[\tanh(\beta h_0)\beta
z_0\right],\nn
\label{P(h0z0)}
\end{eqnarray}
where ${\cal N}$ is a normalization constant.
The self-consistency of
the cavity approach requires that the average correlations on site
$0$
are the same as on the neighboring sites, i.e.,

\begin{eqnarray}
\label{SCcavityQ}
Q^\star&=&\langle m_i^2\rangle = \int dh_0 dz_0 P(h_0,z_0)
\tanh(\beta
h_0)^2,
\\
A^\star&=&\langle m_i \delta m_i \rangle
\\ \nn
&&\hspace*{.5cm}=
\int dh_0 dz_0 P(h_0,z_0)
\frac{\tanh(\beta h_0)\beta z_0}{\cosh^2(\beta h_0)},
\\
\label{SCcavityC}
 C^\star&=&\langle \delta m_i^2 \rangle \\
 \nn &&\hspace*{.5cm}=
 \int dh_0 dz_0 P(h_0,z_0) \left(\frac{\beta z_0}{\cosh^2(\beta
h_0)}\right)^2.
\end{eqnarray}

\subsection{Equivalence between the generalized cavity method
and the two group Ansatz}
The cavity approach with a single reweighting $\exp[-\beta m\Delta F]$ corresponds to an annealed calculation, neglecting correlations and clustering among different states. Let us thus establish its connection with the annealed approximation in the two group formalism.

It is straightforward to convince oneself that the above
self-consistency
conditions Eqs.~(\ref{SCcavityQ}-\ref{SCcavityC})
are identical to the saddle point equations of the
two-group Ansatz (\ref{selfcons2G_1}-\ref{selfcons2G_1C}), evaluated with the help of the annealed free energy expression~(\ref{f:phim_cav}) (see Appendix \ref{app:phi_m} for details). In particular, we find that
 the two group order parameters $Q$, $A$ and $C$  coincide
with the above $Q^\star$, $A^\star$ and $C^\star$.

We finally need to establish the precise correspondence
 between the regularized free energy shift $\Phi_{\rm cav}(m)$
computed within the generalized cavity
method, and the replicated free energy density $\Phi_{\rm 2G}(m)$ computed within
the annealed two-group replica Ansatz. We need to take into
account that by adding a spin to an SK model with $N$ spins, one obtains a system with
slightly stronger couplings (by an average fraction of $1/2N$) than in a standard  system with $N+1$ spins. This is
basically equivalent to raising the inverse temperature to $\beta\rightarrow
 \beta'=\beta(1+1/2N)$ simultaneously with the spin addition.~\cite{MPV86}

We thus expect the relationship
\BEQ
\Phi_{\rm cav}(m)=\Phi_{2G} +\frac{1}{2}\frac{\p(\beta \Phi_{2G})}{\p \beta}
\label{cav_rep}
\EEQ
between the two free energy densities.
Explicit evaluation of the righthand side using Eq.~(\ref{Eq_Phi})
indeed yields
\BEA
&&-\beta m\Phi_{2G}-\frac{\beta}{2}\frac{\partial}{\partial\beta}
(\beta m\Phi_{2G})
\label{phimcavity}
\\
&&\hspace*{3 cm}= \beta \phi_m +m\frac{\beta^2}{2}(1-Q)
-\beta^2 A,
\nn
\EEA
which precisely coincides with $ -\beta m\Phi_{\rm cav}(m)$ from
 Eq.~(\ref{phiofm1}),
if we recall the annealed version of Eq.~(\ref{f:phim_cav}) for $\phi_m$.

We have thus proven the equivalence of the generalized cavity method
and the two group replica calculation for all free energy densities in
the annealed regime $f>f^\star$. Further, we confirmed once more the
interpretation of the order parameters.

To use the cavity method beyond the annealed approximation is a rather
 cumbersome task, see Ref.~[\onlinecite{MPV86}]. It is much easier to
 carry out the two group computation with full replica symmetry
 breaking, even though it is less intuitive than the cavity approach.
 To help the reader  appreciate the physical content of such a
 quenched calculation, we devote the following section
 formalism.

\section{Quenched two group Ansatz: Formalism and Interpretation}
\label{s:quenched}

In Sec.~\ref{ss:2G_ann} we have seen that for free energy densities
$f>f^\star$ the annealed solution of the two group replica approach is
correct. For $f<f^\star$, however, the marginal metastable states are
correlated and thus $Q_{ab},A_{ab},C_{ab}\neq 0$ for $a\neq b$.  In
analogy to the Parisi solution for the ground state of the SK model,
we are looking for a hierarchical breaking of the replica symmetry
encoding the assumption that marginal states are organized in an
ultrametric tree in phase space: the smaller the distance on the tree,
the larger the similarity between the states.  In particular, we
expect the off-diagonal part of the matrices $Q_{ab}$, $A_{ab}$ and
$C_{ab}$ to tend to the functions $q(x)$, $a(x)$ and $c(x)$,
respectively, describing the continuous breaking of replica symmetry.
We continue to denote diagonal entries as $Q_{aa}\equiv Q$,
$A_{aa}\equiv A$ and $C_{aa}\equiv C$.

\subsection{Replicated free energy}
With such an Ansatz for the overlap matrices, the quenched replicated free energy of $m$ copies, Eq.~(\ref{Eq_Phi}), reads
\BEA -\beta m \Phi_{2G}^{\rm qu} &=& \beta\phi(x=0,{\bf y}= (0,0))
\label{Phi2Gfull}
\\
\nn
&&\hspace*{-1cm}
+\frac{\beta^2}{4}m\left(1-Q\right)^2
-\frac{\beta^2}{4} m^2\left[Q^2-\int_0^1dx~q^2(x)\right]
\\
\nn
&&\hspace*{-1cm}-\frac{\beta^2}{2}\left[
2A+A^2+QC-2(1-m)AQ
\right]
\\
\nn
&&\hspace*{-1cm}+\frac{\beta^2}{2}\int_0^1dx\left[a^2(x)+q(x)c(x)+2m~a(x)q(x)\right],
\EEA
where ${\mathbf y} \equiv (y_1,y_2)$. The function $\phi(x,{\bf y})$
is the free energy per replica of a subsystem of $x\cdot m$ coupled
replicas subject to external fields $y_1,y_2$ acting on the two groups
of replicas (representing minima (+) and saddles (-)) with the same
and opposite sign, respectively: $y^\pm = y_1\pm y_2/(2K)$ (cf., the
analogous expression Eq.~(\ref{2G:pheno}) for the
magnetizations). More precisely, we define
\BEA
\label{defofphi}
\exp[x~\phi(x,{\bf y})]&=& \sum_{ s_a^i=\pm 1 }
\exp\left[ {\cal
H}(x,{\bf y},\{s_a^i\})\right],\\
{\cal H}(x,{\bf y},\{s_a^i\})&=&
\frac{\beta^2}{2}\sum_{a,b}^{1,x}\sum_{i,j}^{1,m} s_a^i {\cal
Q}[x]_{ab}^{ij} s_b^j
\nn
\\
&&\hspace{-2cm}+\beta \sum_{a=1}^{x}\left(y_1\sum_{i=1}^m
s_a^i+\frac{y_2}{2K}\left(\sum_{i^+}s_a^{i^+}-\sum_{i^-}
s_a^{i^-}\right)\right). \nn \EEA
Here ${\cal Q}[x]_{ab}^{ij}$ denotes the matrix ${\cal
Q}_{ab}^{ij}-Q^{\sigma_i\sigma_j}(x)$ restricted to a block of
$(xm)
\times (xm)$ replicas.

The representation (\ref{defofphi}) allows for the derivation of
recursion equations for $\phi(x-dx,{\bf y})$ in terms of $\phi(x,{\bf
y})$ (see, e.g., Ref. [\onlinecite{Duplantier}]).  In the limit of
continuous overlap functions ($dx\rightarrow 0$) they reduce to
a Parisi's differential equation
\BEA &&\dot\phi=-\frac{\dot
q}{2} \left[ \frac{\p^2 \phi}{\p y_1^2} +\beta  x\left(\frac{\p
\phi}{\p y_1}\right)^2 \right]
\\
\label{Eqy1y2}
&&-\dot a
\left[
\frac{\p^2\phi}{\p y_1\p y_2}
+\beta x~\frac{\p \phi}{\p y_1}\frac{\p \phi}{\p y_2}
 \right]-\frac{\dot c}{2} \left[ \frac{\p^2
\phi}{\p y_2^2} +\beta x\left(\frac{\p \phi}{\p y_2}\right)^2
\right],\nn
\EEA
where a dot denotes $\p/\p x$.
The boundary condition at $x=1$ follows from the definition (\ref{defofphi}) as
\BEA
\phi(x=1,\mathbf{y})&=&\lim_{K\to\infty}
\frac{1}{\beta}\log \int dh dz ~\mu(h,z)
\\
&&\quad\times\left[2\cosh\beta\left(y_1+h+\frac{y_2+z}{2K}\right)
\right]^{m+K}
\nn
\\
&&\quad\times\left[2\cosh\beta\left(y_1+h-\frac{y_2+z}{2K}\right)
\right]^{-K}
\nn
\\
&
=&\frac{1}{\beta}\log \int dh dz~
\mu(h,z)\left[2\cosh\beta\left(y_1+h\right)
\right]^{m}
\nn
\\
&&\quad\times \exp\left[\beta(y_2+z)\tanh \beta(y_1+h)\right]
\nn
\EEA
where
\BEA
\mu(h,z)~dh~dz&\equiv&\mu({\underline \xi}) d^{2}{\underline \xi}
\\
\nn
&=&
 \frac{d^{2}{\underline \xi}}{2\pi \sqrt{\det {\bf \Delta}}}
\exp\left[-\frac{1}{2}~{\underline \xi}^\dag
\cdot {\bf \Delta}^{-1} \cdot
{\underline \xi}
\right],
\\
{\bf \Delta} &\equiv& \left(\begin{array}{cc} \Delta q \ \ & \ \ \Delta a \\
                                    \Delta a \ \ & \ \
                                    \Delta c\end{array}\right),
\\
{\underline \xi}^\dag &\equiv& (h,z).
\EEA
Here we introduced the notation $\Delta q\equiv Q-q(1)$ etc. for the jump between the diagonal entry and the closest off-diagonal elements of the overlaps. In general, these jumps are finite, indicating that individual marginal states are clearly distinct from their closest neighboring states in phase space. This is in contrast to the situation at $f=f_{\rm eq}$ where the overlap of neighboring states can come arbitrarily close to the self-overlap $Q$.

\subsection{Distribution of local fields $y_1, y_2$}

Following Sommers and Dupont,~\cite{SDJPC84} we also introduce the distribution $P(x,{\mathbf{y}})$ of local fields on the
scale $x$  requiring that
\BEA
\langle s^{i_1}_{a_1} \dots s^{i_r}_{a_r} \rangle&=&\int d{\bf
y} P(x,{\bf y}) \langle \left(s^{i_1}_{a_1} \dots
s^{i_r}_{a_r}\right)\rangle _{{\cal H}(x,{\bf y})}\nn
\EEA
for all
$x\leq\{a_1,\dots,a_r\}\leq 1$. Note that here we need to keep track of the distribution of {\em both} fields $y_1$ and $y_2$.

In the continuous limit, the ensuing recursion relations relating $P$ at $x$ and $x+dx$ lead to the flow equations
\BEA \dot P&=&\frac{\dot q}{2}\left [\frac{\p^2 P}{\p {y_1}^2}-2\beta x~\frac{\p}{\p {y_1}}\left(P \frac{\p \phi}{\p {y_1}}
\right) \right]
\label{eq:P}
\\
&+&\frac{\dot c}{2}\left[\frac{\p^2 P}{\p {y_2}^2}-2\beta x~\frac{\p}{\p {y_2}}\left(P \frac{\p \phi}{\p {y_2}}
\right) \right]
 \nn
\\
&+&\dot a
\left\{\frac{\p^2 P}{\p {y_1}\p {y_2}}-\beta x\left[\frac{\p}{\p {y_1}}\left(P \frac{\p \phi}{\p {y_2}}\right)+\frac{\p}{\p {y_2}}\left(P \frac{\p \phi}{\p {y_1}}
\right) \right]\right\}, \nn
 \EEA
with the boundary condition at $x=0$
 \BEQ
 P(x=0,{\bf y})=\delta^{(2)}({\bf y}).
\EEQ

The joint distribution of local fields $(y_{1},y_{2})$ within a typical
 marginal state (with free energy density $f=f(m)$) is eventually obtained from the convolution

\BEA
P_{\rm qu}({\bf y})&=&\left[2\cosh(\beta y_1)\right]^{m}
 \exp\left[\beta y_2\tanh(\beta y_1)\right]
\label{Pofhquenched2}
\\
\nn
&&\ \ \int  \frac{d^2{\underline \xi}\,e^{-\frac{1}{2}{\underline \xi}^T
\cdot {\bf \Delta}^{-1}\cdot
{\underline \xi}}}{2\pi \sqrt{\det {\bf \Delta}}}
 P\left(1,{\mathbf y}-{\underline \xi}\right)
~e^{-\beta \phi\left(1,{\mathbf y}-{\underline\xi}\right)}
\EEA

The variable transformation (\ref{vartrafo1},\ref{vartrafo2}),
 ${\tilde m}=\tanh(\beta y_1)$ and $\delta m=\beta y_2/\cosh^2(\beta
 y_1)$, yields the joint distribution of local magnetizations and soft
 mode components,
\BEA
\label{Pofmquenched}
P_{\rm qu}({\tilde m},\delta m)&=&\int d^2{\bf y} P_{\rm qu}({\bf y})
\delta({\tilde m}-\tanh(\beta y_1))
\nn
\\
&& \quad
\times\delta\left(\delta m-\frac{\beta y_2}{\cosh^2(\beta
y_1)}\right),
\EEA
one of the central objects of interest
characterizing the metastable states.

\subsection{Self consistency equations for the order parameters}

Using the definitions of $\phi$ and $P$, one can convince oneself
that the continuous limit of the off-diagonal self-consistency equations (\ref{selfcons2G_1}-\ref{selfcons2G_1C}) can be cast into the form
\BEA
q(x)&=&\left[\left(\frac{\p \phi(x,\mathbf{y})}{\p y_2}\right)^2\right]_x,
\label{selfq}
\\
\nn
a(x)&=&\left[\frac{\p \phi(x,\mathbf{y})}{\p y_2}
\left(\frac{\p \phi(x,\mathbf{y})}{\p y_1}-m\frac{\p \phi(x,\mathbf{y})}{\p y_2}\right)
\right]_x,
\\
\label{selfa}
\\
c(x)&=&\left[
\left(\frac{\p \phi(x,\mathbf{y})}{\p y_1}-m\frac{\p \phi(x,\mathbf{y})}{\p y_2}
\right)^2\right]_x,
\label{selfc}
\EEA
where we have introduced the notation
\BEQ
\left[o(\mathbf{y})\right]_x\equiv
\int_{-\infty}^\infty \hspace*{-3mm} dy_1
\int_{-\infty}^\infty \hspace*{-3mm} dy_2
~P(x,\mathbf{y})~o(\mathbf{y})
\EEQ
to denote an average over the local field distribution at the scale $x$.
Alternatively, Eqs.~(\ref{selfq}-\ref{selfc}), can be derived from a variational formulation of the quenched problem.~\cite{CLPR04b}

For the diagonal (intra-state) overlaps $(a=b)$ we have
\BEA
Q&=&\langle {\tilde m}^2\rangle,\\
A&=&\langle {\tilde m} \,\delta m\rangle,\\
C&=&\langle\delta m^2\rangle,
\EEA
where averages are over the joint distribution Eq.~(\ref{Pofmquenched}).
Integrating out explicitly the soft modes, this can be cast into a form similar to the annealed equations
(\ref{saddleQz}-\ref{saddleCz}),
\BEA
&&Q=\left<\tanh^2(\beta y_1)\right>,
\label{quen:saddleQz}
\\
&&A+1-Q= -Q\frac{\Delta a}{\Delta q} -m Q
\label{quen:saddleA2z}
+ \frac{\left<y_1 \tanh\left(\beta y_1\right)\right>}{\beta\Delta q},
\\
\label{quen:saddleCz}
&&C-2A-m (1-Q)=m^2Q+2 mQ\frac{\Delta a}{\Delta q}+Q\left(\frac{\Delta a}{\Delta q}\right)^2
\nn\\
&&\hspace*{1.5 cm}-2\left(m+\frac{\Delta a}{\Delta q}\right)
\frac{\left<y_1\tanh\beta(y_1)\right>}
{\beta \Delta q}
\\
\nn
&&\hspace*{1.5cm}
-\frac{1}{\beta^2 \Delta q}\left(1-\frac{\left<y_1^2\right>}{\Delta q}
\right),\EEA
where $\left<\dots\right>$ denotes an average over $P_{\rm qu}(y_1)\equiv \int dy_2 P_{\rm qu}(y_1,y_2)$.


\section{Internal consistency and Thermodynamic stability}
\label{s:stab}
Sections~\ref{s:2G}, \ref{s:TAPcount} and \ref{s:cav}
provide three equivalent methods to capture the properties of marginal states
at a given  free energy density above $f^\star$, and the previous section extended
the formalism to the low energy regime. Yet, we did not bother so far about
the internal consistency and thermodynamic stability of the obtained solutions.
Moreover, we merely stated the existence of marginal states without analyzing in more detail the eigenvalue
spectrum of the free energy Hessian and the related question of local stability of the states.
In this section we address these issues, and show how to obtain more information about the local environment of the marginal states.

Even though the analysis of the local stability and the thermodynamic
consistency are two a priori very different aspects of the problem, they turn out to be closely related.

\subsection{Internal consistency of the 2G solution}
In order to understand the local free energy landscape of a given
metastable state (TAP solution), we need to characterize the free
energy Hessian of a typical local minimum, or in other words, the
inverse of the susceptibility matrix, $\chi_{ij}^{-1}=\beta \p_i\p_j
F_{\rm tap} (\{m\})$.\cite{Ple0,Ple1} The fluctuation-dissipation
relation requires that the ("zero-field cooled") susceptibility, i.e.,
the trace of $\chi$, be equal to $\beta(1-Q)$.\cite{Ple1} However, not
all solutions of the TAP equations satisfy this constraint, but only
those for which the inequality
\BEA
 && x_{\rm
P}\equiv 1-\beta^2\frac{1}{N} \sum_i (1-m_i^2)^2 \geq 0
\label{plefka}
 \EEA
holds. Other solutions are unphysical.

This condition has always to be checked separately, after having obtained
 a self-consistent solution of site magnetizations $m_i$.

For $N\to \infty$, this can be rewritten as
\BEQ
x_P=1-\beta^2\left(1-2 q +\left<{\tilde m}^4\right>\right) \geq 0
\label{plefkac2}
\EEQ
where the average is over the appropriate magnetization distribution
 (Eq.~(\ref{Pofhann}) with ${\tilde m}\equiv \tanh(\beta h)$ in the annealed case,
and Eq.~(\ref{Pofmquenched}) in the quenched case).

In the regime $f>f^\star$ the condition (\ref{plefkac2}) is satisfied
as a strict inequality in all marginal states. In contrast, if one
aims at describing genuinely stable minima by imposing saddle points
that conserve the BRST symmetry, the condition (\ref{plefkac2}) is
always violated.  This leads to the conclusion that in the SK model
there are no genuinely stable TAP states which are not closely related
to the family of dominating marginal states (see
Ref. [\onlinecite{CLPR03}] [annealed case] and
Refs.~[\onlinecite{CLPR04b,CLPR04a}] [quenched case]). While we know
from the regularization procedure that there are actually stable
states, they are always close to being marginal, sharing similar
properties with the dominating marginal states.

The only thermodynamically consistent states that do not break the
 BRST symmetry are the states at the equilibrium free energy density
 $f_{\rm eq}$ for which the criterion (\ref{plefkac2}) is marginally
 satisfied as an equality.

We note that Eq.~(\ref{plefkac2}) is actually equivalent to the
requirement of a positive {\em replicon} $\Lambda_R$, defined as the
smallest eigenvalue characterizing the fluctuations of the replicated
free energy as a function of the order parameter matrix ${\cal
Q}_{ab}$.  In fact, $\Lambda_R$ turns out to be simply proportional to
$x_P$.\cite{MPV87} The implications of a vanishing replicon, and the
possibility of its simultaneous occurrence with the breaking of the
BRST symmetry in the regime $f_{\rm eq}<f<f^\star$ will be discussed
below.

\subsection{The free energy Hessian}
\label{ss:spectrum}

The extensive part of the spectrum of the inverse susceptibility
matrix ${\mathbf{\chi}}^{-1}(\{m\})$ in a generic TAP state $\{ m_i\}$
was determined in Ref.~[\onlinecite{Ple1}] from an analysis of the TAP
equations, neglecting terms of order $O(1/N)$.  The extensive part
starts off as a semicircle
\BEA
\label{f:spectrum}
\rho(\lambda)&=&\frac{N}{\pi\sqrt{p}}\sqrt{\lambda-\lambda_0},
\hspace*{1 cm}  \lambda-\lambda_0 \ll 1,
\EEA
where the lower band edge $\lambda_0$, together with the resolvent
$r_0=\Tr [(\lambda_0-{\mathbf{\chi}}^{-1})^{-1}]$, follows from the solution of
\BEA
\label{Plefka1}
r_0&=&-f_1(r_0+\lambda_0),\\
\label{Plefka2}
1&=& f_2(r_0+\lambda_0),\\
f_n(x)&\equiv & \frac{1}{N}\sum_{i=1}^N \frac{1}{\left(\frac{1}{\beta(1-m_i^2)} +\beta(1-Q)+x\right)^{n}}.
\EEA
The semicircle's amplitude is controlled by $p=f_3(r_0+\lambda_0)$.
The support of the continuous part was proven to be strictly
 positive,\cite{Ple1} $\lambda_0\geq 0$. However, this result, valid
 to leading order in $N$, does not exclude the presence of a
 sub-extensive number of negative eigenvalues.  In fact, such regions
 of phase space {\em must} exist as guaranteed by the Morse theorem.

One can easily check that if $x_P=0$, the solution of
Eqs.~(\ref{Plefka1},\ref{Plefka2}) is $\lambda_0=0,r_0=-\beta(1-Q)$,
while $\lambda_0=0$ implies $x_P=0$ and $r_0=-\beta(1-Q)$. In
particular, the vanishing of the parameter $x_P$ and of the 'band gap'
$\lambda_0$ occur simultaneously:\cite{BM79,Ple1}
$\lambda_0\simeq x_P^2/(4 p)$.

At finite $N$ the spectrum exhibits tails below the extensive band
 edge $\lambda_0$ which may even extend to negative
 eigenvalues. However, as usually in random matrix problems, it is
 reasonable to assume that their density decays exponentially as
 $\exp(-aN^{1/6})$,\cite{Metha} so that they do not survive for large $N$.

However, there is one eigenvalue below the gap edge that survives in
the thermodynamic limit.  Indeed, including the corrections of order
$O(1/N)$ in the analysis of TAP states, one finds that the Hessian
possesses a single isolated eigenvalue, which is not captured by the
analysis to leading order.  Such an eigenvalue is a common feature of
many mean-field spin-glass models. To our knowledge it was first
encountered as {\em the longitudinal eigenvalue} in the spherical
$p$-spin model\cite{CLRpSP} where it is positive and the corresponding
states are genuine minima.  The situation is presumably similar in the
free energy regime of the $p$-spin Ising model where states are
stable.\cite{CLR05,L05} In contrast, for the regime $f>f^\star$ of the
SK model it was proven that the isolated eigenvalue is exactly zero in
the thermodynamic limit.~\cite{ABM,PRJPA04} This provided the first
evidence for the marginality of the dominant states.

An exactly vanishing eigenvalue should actually be present in any
BRST-breaking states. In this paper we used this insight as a starting
point, assuming a soft eigenvalue $\lambda_{\rm soft}$, and deriving
its self-consistency in various ways in the preceding sections.

\subsection{The spin glass susceptibility}

The spin-glass susceptibility is defined as
\BEA
\label{chiSG}
\chi_{\rm SG}=\frac{1}{N}\sum_{ij}\chi_{ij}^2=
\frac{1}{N} \Tr\,\chi^2=\frac{1}{N}\sum_{j=1}^N \frac{1}{\lambda_j^2},
\EEA
where $\lambda_j$ are the eigenvalues of the Hessian.
It has a simple expression in terms of the above introduced parameter
$x_P$~\cite{BM79}
\begin{equation}
\chi_{\rm SG}=\frac{1-x_P}{x_P},
\end{equation}
which is valid for physical states with $x_P\geq 0$.

As $x_P\to 0$ the spin glass susceptibility diverges. As mentioned above, this happens when the gap $\lambda_0$ vanishes,
 i.e., when there is an accumulation of eigenvalues of the Hessian at
 $\lambda=0$, cf.~Eq.~(\ref{f:spectrum}).  Such states are "fully
 marginal", in the sense that they possess an extensive number of soft
 modes.~\cite{fn5}

In marginal states with only one soft mode and a finite gap
$\lambda_0$ to the continuous part of the spectrum, the spin glass
susceptibility remains finite.  Indeed the linear susceptibility
diverges only along the marginal direction which results in a
non-extensive effect for $\chi_{\rm SG}$.\cite{fn6}

\begin{figure}[th]
\resizebox{0.4\textwidth}{!}{\includegraphics{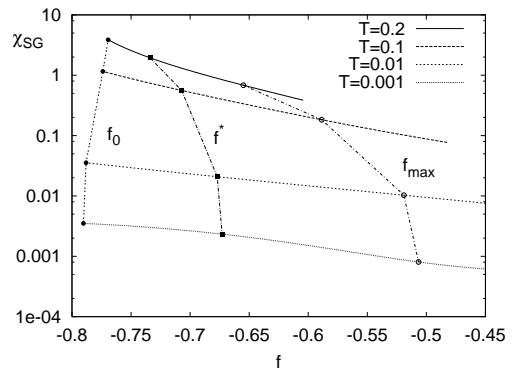}}
\caption{Spin-glass susceptibility of marginal states, $\chi_{\rm SG}$, versus free energy density $f$ at different temperatures
$T=0.2,0.1,.0.01,0.001$. The susceptibility is plotted in log-scale. The physical results holds for $f>f^\star$. $f_0$ is the point at which $\Sigma$ goes to zero in the annealed approximation.
}
\label{fig:chi_sg}
\end{figure}


\subsection{Discussion}
\label{ss:scenarios}
At this stage, we only have an exact description of the free energy
regime $f>f^\star$, while an analysis of the low energy regime down to
$f_{\rm eq}$ requires the quenched solution of the two-group equations
of Sec.~\ref{s:quenched}. The latter is a technically difficult task
and will be carried out elsewhere.~\cite{inprog} Here, we will content
ourselves with a discussion of possible scenarios for the local
environment of dominating states as one decreases the free energy.

The spectrum of the Hessian of a TAP state depends on its free energy
density $f$. As we have seen above, all states with free energies
$f\geq f^\star$ have a finite gap $\lambda_0>0$ in the spectrum, while
the thermodynamically dominant state at $f_{\rm eq}$ is fully
marginal,~\cite{BM79} as implied by the vanishing of the replicon,
$\lambda_0\propto x_P^2\propto \Lambda_R^2 = 0$. The latter is an
interesting situation, which leads to a non-trivial dynamical behavior
owing to the multitude of flat directions in phase space. It is thus
an important question to establish whether full marginality is a
unique property of the equilibrium state, or whether there actually
exists a window of free energies $f_{\rm eq}\leq f\leq f_{\rm gap}$
where the dominant states are fully marginal.  Here $f_{\rm gap}$
denotes the highest free energy density where the gap $\lambda_0$ in
the Hessian spectrum vanishes.

\begin{figure}[th]
\resizebox{0.4\textwidth}{!}{\includegraphics{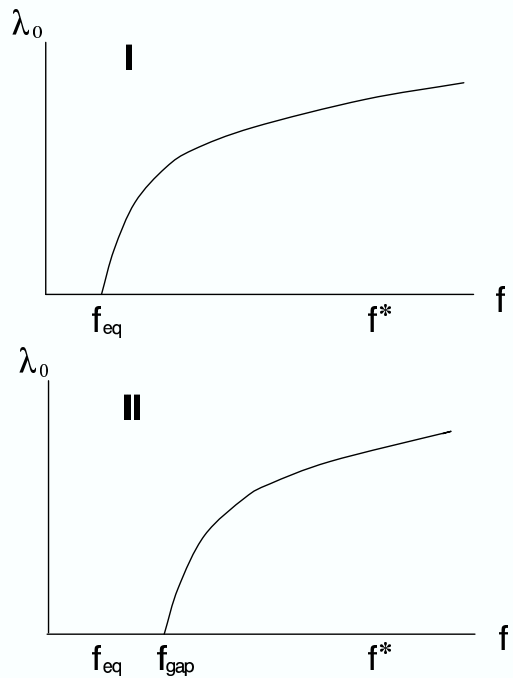}}
\caption{Possible scenarios for the gap in the spectrum,
$\lambda_{0}$, at free energy densities below the threshold of
validity of the annealed approximation, $f^\star$. Scenario I: the gap
$\lambda_0$ goes to zero at $f_{eq}$, as well as the soft mode
overlaps $\{a(x),A\}$ and $\{c(x),C\}$. Scenario 2: the gap goes to
zero at higher free energy ($f_{\rm gap}$) implying the existence of
{\em fully marginal states} also above the global minimum.}
\label{fig:gap}
\end{figure}

The following scenarios are possible in principle:

\begin{itemize}
\item{Scenario I (see Fig. \ref{fig:gap}, top).}  In the simplest
scenario, the gap $\lambda_0$ tends smoothly to zero as $f$ tends to
$f_{\rm eq}$ (together with the BRST breaking order parameters $A$ and
$C$) and thus $f_{\rm gap}=f_{\rm eq}$.  While the soft mode tends to
become orthogonal to the magnetization ($A\to 0$) its amplitude
vanishes simultaneously ($C\to 0$) as the equilibrium state is
approached.  Under this scenario, only the states at $f_{\rm eq}$ are
fully marginal and the formalism developed in
Secs. \ref{s:2G},\ref{s:quenched} should provide a consistent
description for the dominant states (with a finite complexity) at all
free energies and temperatures.

\item{Scenario II (see Fig. \ref{fig:gap}, bottom).}
Fully marginal states already occur
at $f_{\rm gap}>f_{\rm eq}$. In this case, it is a priori not clear whether
 the notion of the (isolated) soft eigenvalue continues to make sense,
since it plunges into a continuum of other eigenvalues.

If, nevertheless, one of the many soft modes can be singled out (e.g.,
with the help of the projector term appearing to order $1/N$ in the
TAP equations\cite{ABM,PRJPA04}), the two group formalism may probably
still describe this regime. Conversely, if one finds a
thermodynamically consistent two group solution with $\lambda_0=0$,
this strongly suggests that the singled out soft mode still preserves
a meaning.  In this case, the solution would still be characterized by
broken BRST symmetry ($C>0$), but we conjecture a vanishing order
parameter $A=0$: while the magnitude of the soft mode stays finite,
its direction is expected to be orthogonal to the magnetization vector
in phase space, as we will argue in the next section.  Further, we
would expect that this branch of solutions continuously joins the BRST
symmetric Parisi solution, (i.e., $C \rightarrow 0$) as $f \rightarrow
f_{\rm eq}$.  Such a branch would yield an extensive complexity,
implying an exponential number of fully marginal states.

On the other hand, the isolated soft mode might lose its meaning
 as the gap closes, and the two group Ansatz might have no solution
 with non-zero soft mode amplitude $C$, suggesting that both $A$ and
 $C$ vanish at $f_{\rm gap}$.  With $A=C=0$, however, the description
 of the system reduces to the standard Parisi ("one group" - BRST
 symmetric) Ansatz, and it has been shown in earlier
 studies~\cite{CLPR04a,CLPR04b} that no thermodynamically consistent
 BRST symmetric continuation of the equilibrium state at $f_{\rm eq}$
 exists, not even within a quenched RSB computation.  The only option
 left would then be the possibility that no states at all exist in the
 interval $]f_{\rm eq},f_{\rm gap}[$, unless a completely different
 replica symmetry breaking scheme is considered.  This last scenario
 is rather unlikely, given that the absence of states between $f_{\rm
 eq}$ and $f_{\rm gap}$ is hardly reconcilable with numerical studies
 of TAP solutions~\cite{CGPnum} which provide evidence for TAP states
 basically at all free energies.

\end{itemize}

\section{Local field distribution in the annealed regime}
\label{s:lowT}

In this section we analyze in more detail the local field distribution
$P_{\rm ann}(h)$, Eq.~(\ref{Pofhann}), of marginal states in the free
energy regime $f\geq f^\star$ where the annealed description applies.
We remind the reader that the distribution depends on the selected
free energy density via the Legendre transform parameter $m$.

\subsection{Low temperature analysis}
The low temperature limit of the complexity was studied long ago in
Refs.~\onlinecite{CDDJP80,BMan,TanakaEdwards}. Here, we focus on the
susceptibility and relate it to the properties of $P(h)$ which are
rather unexpected and, to our knowledge, have not been discussed
before.

As usual, the relevant range of the Legendre parameter $m$ scales as
$m\sim T$ at low temperatures. We thus use the variable $\omega=m/T$
to obtain a sensible $T=0$ limit.  We anticipate that the
susceptibility behaves as $\chi=\beta(1-Q)\sim T$, similarly to what
is known from Parisi's ground state solution. This assumption will
turn out to be self-consistent.  The saddle point equations
(\ref{saddleA2z}-\ref{saddleCz}) for the order parameters $A,C$
suggest a low temperature scaling
\BEQ
\label{lowTAC}
\lim_{\beta \to \infty} \beta A =\lim_{\beta \to \infty} \frac{\beta
C}{2} \equiv-\epsilon-\frac{\omega}{2},
\EEQ
where
\BEA
\epsilon &=& -\frac{1}{2}\lim_{\beta \to \infty}
 \int_{-\infty}^\infty dh~P_{\rm ann}(h)~ h~ \tanh(\beta h)
\nn
\\
&=& -\frac{1}{2}\lim_{\beta \to \infty} \int_{-\infty}^\infty dh~
P_{\rm ann}(h) ~|h|
\label{f:espilon}
\EEA
is the ($\omega$-dependent) energy density of the selected states, as
follows from the low temperature limit of the energy $E =-1/2 \sum
_{ij} s_iJ_{ij} s_j$ and the local fields $h_i = \sum _{j} J_{ij}
s_j$.

In the regime of local fields
\BEA
T \log(1/T) \ll h \ll \beta A \approx -\epsilon-\omega/2,
\EEA the
field distribution Eq.~(\ref{Pofhann})
takes the low temperature form
\begin{eqnarray}
\label{P(h0)simple}
P_{\rm ann}(h) &=&  {\cal N}_0^{-1}
\exp\left[ \frac{\omega}{2}|h|-\frac{(|h|+\epsilon)^2}{2} \right],\\
{\cal N}_0&=& \int d h \exp\left[ \frac{\omega}{2}|h|
-\frac{(|h|+\epsilon)^2}{2} \right].
\EEA
The self-consistency Eq.~(\ref{saddleA2z}) then reduces to\cite{CDDJP80}
\BEA
\label{epsilon}
\epsilon&=&-\frac{1}{2}\int_{-\infty}^\infty d h~
|h|~P_{\rm ann}(h) \\
\nn
&=&\frac{\epsilon}{2}-\frac{\omega}{4}
-\frac{\exp[-\epsilon^2/2]}{{\cal N}_0},
\EEA
which relates $\epsilon$ and the Legendre parameter $\omega$.

\subsection{Soft mode direction}
As a corollary of the low $T$ scaling Eq.~(\ref{lowTAC}) we obtain the behavior
 of the angle between the soft mode and the magnetization as a function of (free) energy density:
\BEA
\label{anglelowT}
\gamma(\epsilon;T)\approx \frac{\pi}{2}-[-\epsilon-\omega(\epsilon)/2]^{1/2}T^{1/2},\quad \epsilon\geq\epsilon^\star.
\EEA
where the $T=0$ limit of the annealed threshold $f^\star$ is $\epsilon^\star= -0.672$.
In particular, we note that the lower the temperature the more the soft mode
 tends to be perpendicular to the magnetization, cf. Fig.~\ref{fig:gamma}.

\subsection{Susceptibility}
\label{s:lowTsusc}
In order to confirm that indeed $\chi\sim T$ we write
\BEA
\label{QSC2}
\chi=\beta(1-Q)&=&\int_{-\infty}^\infty d h P_{\rm ann}(h) \frac{\beta}{\cosh^2(\beta h)}.
\EEA
The integral is dominated by $T\ll h\sim T\log(1/T)$,
where the local field distribution simplifies to
\begin{eqnarray}
P_{\rm ann}(h) &\approx&  {\cal N}_0^{-1}
\exp\left[ - \frac{\beta(-\epsilon-\omega/2)}{\cosh^2(\beta h)}
-\frac{\epsilon^2}{2} \right].
\label{P(h0)smallh}
\end{eqnarray}
Changing variables to  $\xi=\beta/\cosh^2(\beta h)$, we indeed find
\BEA
\label{susc}
\chi &\approx& T\int_{0}^{\beta} d\xi
\frac{\exp[-(\epsilon-\omega/2)\xi-\epsilon^2/2]}{{\cal N}_0}\approx \frac{T}{2},
\EEA
where we have made use of Eq.~(\ref{epsilon}). Surprisingly, the susceptibility
 does not depend on the energy density $\epsilon$ in the low $T$ limit.
The origin of its linear $T$ dependence will be discussed further below.

\subsection{Discussion}

\subsubsection{Pseudogap and full marginality}
The local field distribution Eq.~(\ref{P(h0)smallh}) is rather
peculiar (it is plotted for $T=0.01$ in Fig.~\ref{fig:Py}). It differs
significantly from that in the equilibrium state, both for thermally
active sites (with fields of order $h\sim T$) as for larger fields of
order $h\sim O(1)$.

The field distribution of the ground state
 is known to assume a universal low temperature scaling form~\cite{SDJPC84}
\BEA
\label{GSscaling}
P(h,T)&=&\left\{\begin{array}{cc} T\psi(h/T) & \textrm{for  } h\sim T,\\
{\rm cst.}\, |h| & \textrm{for  } h\gg T,
\end{array}\right.
\EEA
which is related to an asymptotic fixed point in Parisi's flow
equations.~\cite{Pankov06} The expression (\ref{GSscaling}) implies a
susceptibility linear in $T$, arising from $O(T^2)$ spins with a local
linear response of order $1/T$.

In contrast, in the marginal states at higher energies, $f>f^\star$,
the linear susceptibility, Eq.~(\ref{susc}), is due to $O(T)$ spins
with a local response of order $O(1)$ (on sites with fields $h\sim
T\log(1/T)$). On the other hand, the density of really active spins
(subject to local fields of order $h\lesssim T$) is exponentially
suppressed at low temperature, \BEA
\label{Pat0}
P_{\rm ann}(h=0)\sim \exp[-\beta(-\epsilon-\omega/2)].
\EEA

The scaling behavior Eq.~(\ref{GSscaling}) in the ground state is closely related to its full marginality.~\cite{TAP,SDJPC84}
Indeed, full marginality requires the vanishing of the replicon - see Sec. \ref{ss:spectrum}
($\Lambda_R\propto x_P\to 0$),
which in turn is equivalent to the marginality condition
\BEA
\label{GSmarginality}
1=\beta^2\int d h \frac{P_{\rm qu}(h)}{\cosh^4(\beta h)}.  \EEA Here,
the local field distribution $P_{\rm qu}(h)$ is given by the quenched
Parisi solution (i.e., Eq.~(\ref{Pofhquenched2}) in the limit
$A=a(x)=C=c(x)=\Delta q=0$, with $y_1\equiv h$ and $y_2$ being
trivially integrated out).  From the equality (\ref{GSmarginality}),
valid for all $T$, one infers the low temperature scaling
(\ref{GSscaling}).

With the connection between full marginality and $h/T$-scaling in
mind, the lack of the latter in higher marginal states (with only one
soft mode) may not be too surprising, since we know that above
$f^\star$ the states are not fully marginal (the replicon $\Lambda_R$
remains finite throughout).  Further consequences for the regime $h=
O(1)$, such as the absence of the linear pseudogap implied by
Eq. (\ref{GSscaling}), will be discussed in the next subsection.

There is a noteworthy aspect of the nearly hard gap in the low field
 distribution Eq.~(\ref{Pat0}): the exponential suppression of
 $P_{\rm ann}(h\lesssim T)$ (see inset of Fig. \ref{fig:Py}) decreases with
 $-\epsilon-\omega/2 \approx \beta A$ and eventually would vanish at
 low enough energies as $A \rightarrow 0$. This is, however, preempted
 by the breakdown of the annealed approximation and the occurrence of
 full replica symmetry breaking at $f^\star$ which requires a quenched
 computation.  Nevertheless, the above observation suggests that the
 vanishing of $A$ is connected with the onset of {\em full}
 marginality, i.e., the approach to zero of the continuous spectrum of
 the free energy Hessian. This is supported by a result by Parisi and
 Rizzo~\cite{PRJPA04} who showed that as the continuous spectrum
 approaches $\lambda_0 \rightarrow 0$, the (isolated) marginal mode
 becomes orthogonal to the magnetization (that is, $A=0$ in our
 formalism).  We thus conjecture that in general the vanishing of the
 gap in the spectrum, $\lambda_0\propto x_P^2=0$, implies the
 vanishing of $A_{ab}$. With respect to the discussion of the previous
 Sec.~\ref{ss:scenarios}, this is guaranteed to happen in scenario I,
 while it constitutes a non-trivial prediction for scenario II.

\subsubsection{On stability and slow dynamics}
 The field distribution Eq.~(\ref{P(h0)smallh}) differs from the
equilibrium one also in the range of fields $h=O(1)$. In the limit
$T\rightarrow 0$, the fraction of fields with a fixed magnitude $T\ll
h\ll 1$ remains bounded from below by a finite constant,
cf. Eq.~(\ref{Pofhann}). In particular, there is {\em no} linear
pseudogap $P(h)\sim |h|$, in contrast to the equilibrium state,
cf. Eq.~(\ref{GSscaling}). Strictly at $T=0$ there is a finite
probability of configurations at $h=0$, while at finite $T$ there is
an almost hard gap on the scale $T\log T$ as discussed in the previous
section.

\begin{figure}[tbp]
\resizebox{0.4\textwidth}{!}{\includegraphics{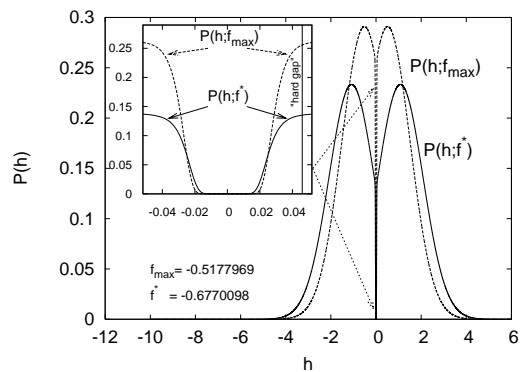}}
\caption{Local field distribution Eq.~(\ref{P(h0)smallh}) in the
annealed regime for $f_{\rm max}$ and $f^\star$ at $T=0.01$.  In the
inset the region around $h=0$ is enlarged to show the nearly hard gap
occurring at any $f$ in the annealed regime.  The vertical line marks
$T\log T$ which is the scale $h_{\rm gap}$ at which the gap is
expected to open.  }
\label{fig:Py}
\end{figure}

The absence of a linear pseudogap immediately raises the question
 about the stability of these states.  As was realized in the early
 days of the SK model, a linear pseudogap is the minimal suppression
 of the low field distribution for a truly stable state (see, e.g.,
 the discussion in Ref. [\onlinecite{Anderson78})], a configuration with
 a finite density of local fields around $h=0$ being unstable to the
 flipping of a finite number of spins (exactly at $T=0$). A very
 similar argument lead Efros and Shklovskii to infer the presence of
 the Coulomb gap in long range interacting electron
 glasses.~\cite{ES75} Recently, it was recognized that both pseudogaps
 are related to the full marginality of the equilibrium
 state.~\cite{SDJPC84,MuellerIoffe04,MuellerPankov06}

The absence of a pseudogap in marginal states would seem to render
 them unstable with respect to a finite number of spin flips.  While
 this is true strictly at $T=0$,\cite{BMan} the finite temperature
 analysis is more subtle, because the thermodynamic limit and the
 $T\to 0$ limit do not commute.  One can apply  stability arguments
 analogous to those of Ref.~[\onlinecite{Anderson78}] to the field
 distribution Eq.~(\ref{P(h0)smallh}) of marginal states above
 $f^\star$. Such an analysis shows that at low but finite temperature
 a collective flip of order $O([T\log(1/T)]^2N)$ spins (randomly
 chosen among the sites with small local fields) is necessary in
 general to render the state unstable.  This contrasts with $O(T^2 N)$
 flips for states with a linear pseudogap, but no hard gap on the
 scale of $h\sim T$.  The logarithmic enhancement of stability against
 {\em random} spin flips for $f>f^\star$ is indeed due to the presence
 of an almost hard gap on the scale of $h_{\rm gap}\sim T\log(1/T)$,
 see inset of Fig.~\ref{fig:Py}. One may therefore expect that a finite $T$
 dynamics based on random activated spin flips exhibits similarly long
 (if not longer) escape times as low energy states with linear
 pseudogap.

On the other hand it is clear that a collective flip of a set of spins
 with a significant projection onto the marginal mode will take the
 system immediately out of the local state since the free energy
 barrier presumably decreases to zero in the thermodynamic limit as
 suggested by recent simulations.\cite{ABBM} However, such an escape
 may be very difficult to realize since it amounts to flipping a large
 number of spins in a concerted manner.  In summary, the presence of a
 single marginal direction does not seem to decrease the metastability
 of higher-lying marginal states in a dramatic way, rather their
 stability seems comparable to that of states closer to equilibrium.

At the present stage the question as to the consequences of marginal
states and their local environment on the dynamics is still
open. Nevertheless, assuming that the dynamics is eventually dominated
by the 'easy' escape via the marginal mode, we may expect that the
rate at which the Edwards-Anderson parameter $Q$ decreases depends
crucially on the angle $\gamma$ between magnetization and soft mode,
cf., Eq.~(\ref{f:gamma}) and slows down as
$\cos(\gamma)=A/\sqrt{CQ}\rightarrow 0$ with decreasing $f$. This is
illustrated in Fig.~\ref{fig:Q_gamma} where we plot the decreasing
overlap $Q$ as a function of the increasing angle $\gamma$.

\begin{figure}[t!]
\resizebox{0.4\textwidth}{!}{\includegraphics{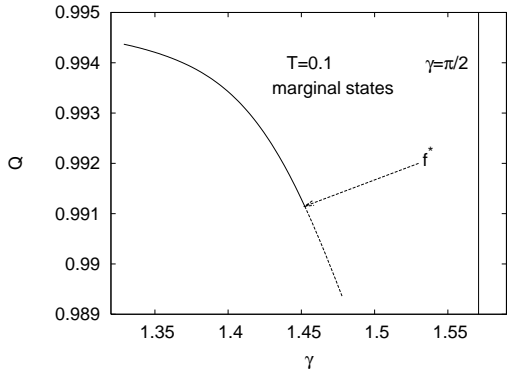}}
\caption{Overlap $Q$ versus angle $\gamma$ a $T=0.1$.  The soft mode
becomes more and more orthogonal to the magnetization vector as the
free energy $f$ and the Edwards-Anderson parameter $Q$ decrease.  The
vertical line indicates $\gamma = \pi/2$, which is not reached within
the annealed regime.  The dashed low energy part of the curve
$Q(\gamma)$ is unphysical in the annealed approximation.  }
\label{fig:Q_gamma}
\end{figure}

This picture is consistent with relaxation dynamics that takes the
 system towards lower and lower lying metastable states. Indeed,
 $Q(f)$ {\em decreases} as $f$ decreases (at least in the annealed
 regime, see Fig.~\ref{fig:Q_f}). At first sight, this result looks
 counterintuitive, being opposite to the behavior of the self-overlap
 in familiar one step glasses with genuinely stable states (e.g.,
 $p$-spin models below the marginality threshold) where $Q(f)$
 increases as one descends to lower energies.  It may be interesting
 to note that an increase of $Q(f)$ with decreasing $f$ is found in
 the BRST-symmetric solution of Eqs. (\ref{f:Qab_sc}-\ref{f:Cab_sc}),
 see inset of Fig. \ref{fig:Q_f}. However, in the SK model this
 solution is thermodynamically inconsistent at all free energies above
 $f_{\rm eq}$, even within a quenched computation.\cite{CLPR04b}

\begin{figure}[t!]
\resizebox{0.4\textwidth}{!}{\includegraphics{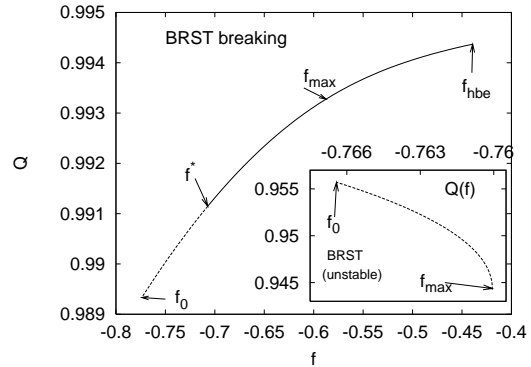}}
\caption{Free energy dependence of the order parameter $Q$ in the two
group approach at $T=0.1$. Note the {\em decrease} of $Q(f)$ with free
energy.  For comparison, we also plot the (unphysical) BRST-symmetric
solution where $Q(f)$ displays the opposite behavior. $f_{\rm max}$
denotes the value of maximum $\Sigma$, $f_0$ and $f_{\rm hbe}$ the
point of vanishing $\Sigma$ (lower band edge and upper band edge,
respectively). Above $f^\star$ the annealed solution is exact.  }
\label{fig:Q_f}
\end{figure}

 The above observation suggests to think of the marginal metastable
states discussed here as relatively small pockets of phase space that
restrict the local magnetizations to slightly {\em higher} values than
in more equilibrated states. Relaxation dynamics out of these shallow
traps, and subsequent descent in free energy will allow the spins to
explore more degrees of freedom and to lower the self-overlap $Q$.
Clearly, it will be important to check such a picture in analytic
studies of glassy dynamics as well as in simulations, for which the
interpretation of the order parameters in terms of soft mode
correlations may provide helpful guidelines.

\section{Outlook and open questions}
\label{s:outl}
Our study of the local landscape around a given metastable state
 leaves open the question as to the organization of these states
 within phase space. In particular, one would like to understand the
 typical fate of a system once it manages to escape from a local
 marginal trap via the soft mode. This would be an important element
 to a more complete understanding of relaxation dynamics as well as to
 avalanche-like dynamics observed when the SK glass is driven with an
 external field.~\cite{Pazmandi99}

 The toy model of Section~\ref{s:toy} was very helpful to understand
the meaning of the two group Ansatz.  Similar models may be guidelines
as to how to use 'vectorial symmetry breaking' in order to obtain
physical information in complicated glasses, and to address questions
such as the above.  A natural example is the random field Ising model,
where an extension of the two group Ansatz to three groups can be used
to describe dominant Griffith singularities, and to capture rare
disorder configurations which describe the emergence of metastable
states in the ferromagnetic regime.~\cite{MullerSilva06}

\subsection{Describing barriers with replicas}
Another interesting application of the two group Ansatz might be the
study of barriers.  Indeed, from the solution of the toy model it is
clear that replica saddle points with finite $k_2\equiv K$ describe
disorder realizations with two stable minima separated by a finite
barrier of order $O(1/K)$. It may be interesting to pursue this idea
and to work at fixed $K<\infty$ in order to impose a certain barrier
height, similarly as fixing $m$ allows one to choose the mean free
energy density of the minimum-saddle pairs. One may hope that such an
approach will yield information on the properties of barriers between
metastable states, which is a notoriously difficult and unresolved
problem in glasses.

\subsection{Properties of states in the correlated regime}
Even though we are able to describe some key features of metastable
states in the high energy regime $f>f^\star$ where most marginal
states are uncorrelated among each other, the analysis of lower-lying
states in the free energy range $f_{\rm eq}<f<f^\star$ requires full
replica symmetry breaking on top of a two-group replica structure. The
assumption that marginal states are clustered in a hierarchical manner
in much the same way as states close to the ground state energy
suggests to look for a Parisi-type Ansatz in this energy range, the
physical content of which we have described in Sec.~\ref{s:quenched}.

It will be important to establish how the characteristics of the local
landscape (marginal mode, gap in the spectrum of the Hessian) behave
as a function of free energy in this regime.  In particular, one would
like to know whether the breaking of BRST-symmetry persists in the
whole interval, as actually suggested by the impossibility to
construct a physical BRST-symmetric solution at any $f>f_{\rm
eq}$.\cite{CLPR04a,CLPR04b} This issue is closely related to the
interesting open question: do any typical states with $f>f_{\rm eq}$
display full marginality, or is this intriguing property (implying a
divergent spin glass susceptibility, critical fluctuations and
dynamics) unique to the ground state?  This important issue will be
clarified by the quenched solution of the two group
Ansatz.\cite{inprog}

It will be interesting to extend the two group analysis to other
mean-field like models, such as random manifolds with a large number
of transverse dimensions~\cite{MPrandomMF} or electron
glasses.~\cite{MuellerIoffe04,MuellerPankov06} In the latter problem,
the presented techniques may help to address experimentally relevant
questions concerning the evolution of the Coulomb gap with decreasing
free energy.

\subsection{Ground state properties in general mean field glasses}
The SK model is rather special in that its equilibrium state is always
fully marginal below the glass transition. This observation, and the
above discussion of the approach to full marginality as a function of
free energy raises a question about the nature of low energy states in
more general mean field glasses.  In the Ising $p$-spin model it has
been established~\cite{CLR05} that above the so-called Gardner
temperature $T_G$ the states in an energy interval $[f_{\rm eq},f_G]$
including the ground state are stable minima described by (one group)
one-step RSB. At higher energies, marginal minima-saddle pairs
dominate, like in the SK model (two-group one-step RSB).
Precisely at $f=f_G$ where the two
regimes meet, the states are fully marginal.  At $T=T_G$ the energy
interval of stable states shrinks to zero, $f_G\rightarrow f_{\rm
eq}$, and consequently the (one step) ground state displays an
instability. What happens to the ground state at lower temperature is
not known. Both the scenario of a permanently fully marginal ground
state similarly to the SK model, or a marginal ground state with a
single soft mode (continuing the branch of solutions at higher free
energies above $T_G$) can be imagined. This open problem is currently
under investigation. Similar questions arise at phase transitions in
many optimization problems (at $T=0$) on diluted lattices. In this
context, it would be very interesting to investigate the physical
meaning of soft modes at zero temperature.\cite{PRPRB05}

\section{Conclusion}
\label{s:conc}
In this paper we have discussed three equivalent approaches to the
description of metastable states in mean field glasses - the replica
two group formalism, the counting of TAP solutions, and the cavity
method. We have focused on the generalizations of these techniques
allowing one to capture the marginal states that generically
dominate the free energy landscape at high enough free energy
densities, and probably play an important role in the dynamics of
these glasses. We found the physical meaning of the additional order
parameters (generalized overlaps) arising in the marginal case: they
describe correlations and relative orientations between the
magnetization and the soft mode of marginal states.

We show the two-group replica formalism to be an effective and compact
computational tool to obtain most of the key features characterizing
the local landscape of marginal states (such as the distribution of
local fields, the spectrum of the free energy Hessian, the direction
of the soft mode, etc.), in particular when it comes to the
description of correlated low energy states. We have made an effort to
exhibit the physical content of this Ansatz in order to render this
tool accessible to a wider audience. Extensions of this Ansatz can be
of use in other problems, e.g., for the study of barriers in
glasses. Simple toy models as introduced in the beginning
(Sec. \ref{s:toy}) may serve as useful guides in the development and
interpretation of such techniques.

Revisiting the uncorrelated high energy regime of metastable states
where an annealed disorder average can be performed, we have discussed
their properties in the light of the new interpretation of order
parameters. Further investigations will be required to establish the
dynamical significance of these states.

\acknowledgments{The authors are indebted to C. De
Dominicis, M. M{\'ezard} and G. Parisi for stimulating discussions.}

\appendix

\section{Expressions for $\phi_m$}
\label{app:phi_m}

The log-trace term (\ref{f:logtr}) in $-\beta m\Phi_{\rm 2G}$ is given by
\BEQ
\label{app_phi_m}
\beta \phi_m \equiv \frac{1}{n}\log\sum_{\{s_a^i\}}
\exp\left(\frac{\beta^2}{2}\sum_{ab}\sum_{ij}s_a^i
{\cal Q}_{ab}^{ij}
s_b^j\right)
\nn
\EEQ
Here we derive two representations of this term, (Eq.~(\ref{f:phim_pp}) and Eq.~(\ref{f:phim_cav})), in order
to facilitate the connection
with the counting of TAP states of Sec.~\ref{s:TAPcount} and the generalized cavity approach of Sec.~\ref{s:cav}, respectively.

Decoupling spin products in Eq.~(\ref{app_phi_m})
by means of Hubbard-Stratonovich transformations, one  obtains the representation 
\BEA
\label{phi0_1}
e^{n\beta\phi_m}&=& K^n\sqrt{ \frac{ \det \Gamma_+\det\Gamma_-}{\det Q}}
\int \prod_{a=1}^n \frac{dg_a}{\sqrt{2\pi}}
\frac{dg_a^+}{\sqrt{2\pi}}\frac{dg_a^-}{\sqrt{2\pi}}
\nn
\\
&&
\sum_{s_a^{i_\pm}}\exp\left[\sum_a\left(\sum_{i_+=1}^{m+K} \beta (g_a+g_a^+)s_a^{i_+} \right.\right.\\
&&\quad\quad\quad \left.\left.+ \sum_{i_-=m+K+1}^{m} \beta (g_a+g_a^-)s_a^{i_-}\right)\right]
\nn
\\
\nn
&&
\hspace{-.8cm} \times  \exp\left[-\frac{1}{2}\left( g_a Q^{-1}_{ab} g_b
+  g^+_a \Gamma_{+,ab}^{-1} g^+_b
+ g^-_a \Gamma_{-,ab}^{-1} g^-_b \right)\right],
\end{eqnarray}

with $n\times n$ matrices
\BEQ
\Gamma_{\pm,ab}\equiv \pm \frac{A_{ab}}{K} +\frac{C_{ab}}{2K^2}.
\label{gamma}
\EEQ

\subsection{Saddle point for $g_a^\pm$ for large $K$.}

For large $K$ one can expand the $\Gamma$ matrices  in Eq.~(\ref{gamma})
as
\BEA
&&\Gamma^{-1}_{\pm} \simeq K\left(\pm A^{-1}+\frac{R}{K}\right),
\\
&&R=-\frac{1}{2}A^{-1}CA^{-1}.
\EEA

We isolate the terms of order $O(K)$ in the exponential of Eq.~(\ref{phi0_1}), and perform the spin sums,
\begin{eqnarray}
\label{phi0_PP}
&&e^{n\beta\phi_m}\simeq
\frac{ |K|^n}{|\det A|\sqrt{\det Q}}
\int \prod_{a=1}^n \frac{dg_a}{\sqrt{2\pi}}
\frac{dg_a^+}{\sqrt{2\pi}}\frac{dg_a^-}{\sqrt{2\pi}}
\\
&&\ \ \  \exp\Biggl\{ m \sum_a \ln 2 \cosh[ \beta(g_a+g_a^+)]\nonumber\\
&&\hspace*{1 cm}+K \biggl[\sum_a \ln \frac{\cosh[\beta (g_a+g_a^+)]}
{\cosh [\beta (g_a+g_a^-)]}
\nn
\\
&&\hspace*{3cm}-\frac{1}{2}\sum_{ab} A^{-1}_{ab}
(g^+_a  g^+_b- g^-_a  g^-_b) \biggr]
 \nonumber\\
&&\hspace*{1.2 cm}-\frac{1}{2}\sum_{ab}\biggl[ g_a Q^{-1}_{ab} g_b
+ R_{ab}\left(  g^+_a g^+_b
+ g^-_ag^-_b \right)\biggr] \Biggr\}.
\nn
\end{eqnarray}
In the limit $K\rightarrow \infty$ we can take the saddle point
approximation for the term in the exponent
proportional to $K$ which leads to the
saddle point equations (with respect to $g_a^+,g_a^-$)
\begin{eqnarray}
\label{SPeqs}
 \tanh[\beta(g_a+g_a^+)]&=&  -\beta^{-1} A^{-1}_{ab} g_b^+,\\
 \tanh[\beta(g_a+g_a^-)]&=&  -\beta^{-1} A^{-1}_{ab} g_b^-,
\end{eqnarray}
determining the location of the saddle point as $g_a^+=g_a^-=g_a^*(\{g_c\})$; $g_a^+$ and
$g_a^-$ can then be integrated out. Decoupling the term $g_a
Q^{-1}_{ab} g_b$ with a further Hubbard Stratonovich field $x_a$, and
changing variables from $g_a$ to
\BEQ
m_a\equiv -\beta^{-1} A^{-1}_{ab} g_b^*(\{g_a\}),
\EEQ
one eventually obtains the expression [Eq.~(\ref{f:phim_pp})],
\begin{eqnarray}
&&e^{n\beta\phi_m}=
2^{nm}
{\int_{-\mbox{\scriptsize i}\infty}^{\mbox{\scriptsize i}\infty}}
\prod_{a}\frac{dx_a}{2\pi{\rm i}}\int_{-1}^{1}\prod_{a}\frac{dm_a}{(1-m_a^2)}
\\
\nn
&&\times\exp\left\{-\sum_{a} \left[x_a\tanh^{-1}m_a
+\frac{m}{2}\log(1-m_a^2)\right]\right.
\\
\nn
&&\hspace*{.5cm}\left.+\beta^2 \sum_{ab}\frac{1}{2}
x_a Q_{ab} x_b+\frac{1}{2}m_aC_{ab} m_b+m_aA_{ab} x_b\right\},
\end{eqnarray}
Note that the sum over $(a,b)$ also includes diagonal terms with $a=b$.
The Gaussian integral over $x_a$ can
 be carried out in principle. For the annealed solution this leads to an expression as in Eq.~(\ref{f:mu(h)}) upon a
change of variables $m_a = \tanh(\beta h_a)$.


\subsection{Cumulant expansion in $g_a^\pm$.}
\label{ss:cumulant}
Alternatively, we may proceed from Eq.~(\ref{phi0_1}) by changing integration variables to
\BEA
\label{vch1}
&& l_a \equiv \frac{g^+_a+g^-_a}{2},
\\
\label{vch2}
&&h_a \equiv g_a + \frac{g_a^++g_a^-}{2},
\\
\label{vch3}
&&z_a \equiv K(g_a^+-g_a^-).
\EEA
The fields acting on the spins in Eq.~(\ref{phi0_1}) turn into
$g_a+g_a^{\pm} = h_a\pm z_a/2K$, while $l_a$ only occurs in the
Gaussian weight and can be integrated out.  The log-trace term then
takes the simpler form
\begin{eqnarray}
\label{app_phim_cav1}
e^{n\beta\phi_m}&=&\frac{1}{\sqrt{\det{\cal M}}}
\int \prod_{a=1}^n\frac{ dh_a dz_a}{2\pi}
\\
\nn
&&\sum_{s_a^{i_\pm}}\exp\left[\sum_a\left(\sum_{i_+=1}^{m+K} \beta (h_a+z_a/2K)s_a^{i_+} \right.\right.\\
&&\quad\quad\quad \left.\left.+ \sum_{i_-=m+K+1}^{m} \beta
(h_a-z_a/2K)s_a^{i_-}\right)\right]
\nn
\\
\nn
&&
\times \exp\left[-\frac{1}{2}\sum_{ab}
{\underline{\xi}}_a^\dagger \cdot {\cal M}^{-1}_{ab}
\cdot {\underline{\xi}}_b\right],
\end{eqnarray}
where ${\underline{\xi}}_a
= (h_a,z_a)$, $a=1,\ldots,n$.
The covariance
matrix ${\cal M}$ is a $2n\times 2n$ matrix given by
\begin{equation}
{\mathbf{\cal M}}=\Bigl(\overbrace{
\begin{array}{c}
 \mathbf{Q} \\
 \mathbf{A}\end{array}
}^n
 \overbrace{
\\
\begin{array}{c}
\mathbf{A} \\
\mathbf{C}
\end{array}
}^n
\Bigr).
\end{equation}
i.e., a $2\times 2$ matrix (as in the annealed case
where off diagonal terms vanish)
with $n \times n$ matrices as entries.
Its determinant is that of
$\mathbf{D}\equiv\mathbf {QC}-\mathbf{A}^2 $, and its inverse is
\begin{equation}
\left[{\mathbf{\cal M}}^{-1}\right]_{ab}=
\sum_{c}\left[D^{-1}\right]_{ac}
\left(\begin{array}{cc}
 \  C_{cb} \ &  \
-A_{cb} \
\\  \
-A_{cb}  \ &  \
Q_{cb}\end{array} \ \ \right).
\end{equation}

Carrying out the spin sums and taking the limit $K\rightarrow \infty$,
one finds the contribution to the replicated free energy
[Eq.~(\ref{f:phim_cav})]

\begin{eqnarray}
\label{app_phim_cav}
e^{n\beta\phi_m}&=&\frac{1}{\sqrt{\det{\cal M}}}
\int \prod_{a=1}^n\frac{ dh_a dz_a}{2\pi}
\\
\nn
&&\exp\left\{
\sum_a \left[m \log 2\cosh(\beta h_a) +
\tanh(\beta h_a) \beta z_a\right]\right\}
\\
\nn
&&
\exp\left[-\frac{1}{2}\sum_{ab}{\underline{\xi}}_a^\dagger
 \cdot {\cal M}^{-1}_{ab}
\cdot {\underline{\xi}}_b\right].
\end{eqnarray}

The integration over $z_a$ in the annealed version of
Eq.~(\ref{app_phim_cav}) leads again to the expression
Eq.~(\ref{f:mu(h)}).

\subsection{Selfconsistency equations and equivalence between two group and cavity approach}
With the help of the measure in Eq.~(\ref{app_phim_cav1}), it is easy
to see that every spin $s^{i_\sigma}$ under the averages in the
two-group self-consistency equations,
Eqs.~(\ref{selfcons2G_1}-\ref{selfcons2G_1C}), is replaced by
$\tanh(\beta(h_a+\sigma\, z_a/2K))$.  In particular, in the limit
$K\rightarrow \infty$ we have
\BEA
\label{app:vartrafo1}
&&s_a^{i_\sigma} \rightarrow \tanh(\beta(h_a+\sigma\,z_a/2K))\\
&&\hspace*{1.cm} \stackrel{K\to\infty}{\rightarrow} \tanh(\beta h_a)\equiv \tilde{m}_a,
\nn\\
\label{app:vartrafo2}
&&
K(s_a^{i_+}-s_a^{j_-})\rightarrow \nn\\
&&\hspace*{.5cm}K\left[\tanh(\beta(h_a+z_a/2K))-\tanh(\beta(h_a-z_a/2K)))\right]
\nn\\
&&\hspace*{1.cm} \stackrel{K\to\infty}{\rightarrow}\frac{\beta z_a}{\cosh^2(\beta h_a)}\equiv \delta m_a,
\EEA
where we introduced the variables $\tilde{m}_a$ and $\delta m_a$. This
leads to the self-consistency equations
(\ref{SCsimpleQ}-\ref{SCsimpleC}), and establishes the equivalence
with the expressions from the cavity approach,
Eqs.~(\ref{SCcavityQ}-\ref{SCcavityC}), on the annealed level.


\section{Generalization of Ward Identities}
\label{app:ward}

In this appendix we derive generalizations of the BRST Ward
identities, reproducing the saddle point equations for marginal
states. We consider sums over TAP states such as in
Eq.~(\ref{f:regul}). We can assume the states to be minima (due to
regularization), which we regard as functions of external fields
$\{h_k\}$, in the sense that $m_\alpha(\{h_k\})$ is the unique
solution of $\partial_i F_{\rm TAP}(\{m_j\}) =h_i$ in the vicinity of
the unperturbed state $m_\alpha(\{h_i=0\})$.

After replicating the system $n$ times (with $n \to 0$ eventually),
 we consider the
identity
\BEA
&&\beta \langle m_i^a x_k^b \rangle {\cal Z}_J^n  = \sum_{\alpha=1}^{{\cal N}_{\rm sol}}
\frac{\partial}{\partial h^b_k} \Bigl\{m_{i,\alpha}^a
\exp \sum_{c=1}^n\Bigl[ \lambda_X X(\{m_{j,\alpha}^c\})
\nn \\
&&\hspace*{3.5cm}
-\beta
m F_{\rm TAP}(\{m_{j,\alpha}^c\})\Bigr]\Bigr\},
\EEA
where the limit $\lambda_X\rightarrow 0$ will be taken at the end.
The function ${\cal Z}_J$ is the TAP partition sum which can be expressed both as the functional integral Eq.~(\ref{action}),
 including the regularizing term $\lambda_X X(\{m_i\})$ in the exponential, or as the sum over TAP states Eq.~(\ref{f:regul}).
The average on the left hand side is taken over the measure defined by the action Eq.~(\ref{action}).
Evaluating the right hand side we find
\BEA
\label{app_Ward1}
&&\beta \langle m_i^a x_k^b \rangle =
\frac{1}{{\cal Z}_J^n}
\\
\nn
&&\times \sum_{\alpha=1}^{{\cal N}_{\rm sol}}
\left(\frac{\partial m_{i,\alpha}^a}{\partial h^b_k}
 +m_{i,\alpha}^a \frac{\partial\lambda_X X}{\partial h^b_k}
 -\beta m m_{i,\alpha}^a \frac{\partial F_{\rm TAP}}{\partial h^b_k}
 \right)  \nn\\
&&\times
\exp\left\{
\sum_{c=1}^n \left[-\beta m F_{\rm TAP}(\{m_{j,\alpha}^c\})
+\lambda_X X(\{m_{j,\alpha}^c\})\right]
\right\} \nn \\
 &&\quad= \langle \delta_{ab}\chi_{ik}^{ab} +m_{i,\alpha}^a
\frac{\partial\lambda_X X }{\partial h^b_k} \rangle\nn
\EEA
where the stationarity of $ F_{\rm TAP}$ was used.
Using the definition of the soft mode, Eq.~(\ref{f:soft_mode}),
and summing Eq.~(\ref{app_Ward1}) over $i=k$, we
establish the generalization of the first Ward identity [Eq.~(\ref{f:Ward1})],
\BEA
\beta \langle m^a x^b\rangle
&=& \delta_{ab} \langle \psi_a \overline{{\psi}_b}\rangle + \beta
\langle m_a \delta m_b \rangle.
\label{app_Ward12}
\EEA

The second identity follows similarly from
the relation
\BEA
&&\beta^2 \langle x_a x_b \rangle {\cal Z}_J^n= \frac{1}{N} \sum_{i,\alpha}
\frac{\partial^2}{\partial h^a_i\partial h^b_i}
\label{app:Ward_gen2}\\
&&\hspace*{.5cm}
\exp\left\{\sum_{c=1}^n [-\beta m F_{\rm TAP}(\{m_{j,\alpha}^c\})
+\lambda_X X(\{m_{j,\alpha}^c\})]\right\}
\nn\\
&&\hspace*{.3cm}= \frac{1}{N} \sum_{i,\alpha}
\left( - m \beta \chi_{ii}\delta_{ab} +\beta^2 \delta
m_{i,\alpha}^a \delta m_{i,\alpha}^b +\frac{\partial^2\lambda_X X}{\partial
h^a_i\partial h^b_i}\right)
\nn\\
&&\hspace*{.4cm}\times
\exp\left\{\sum_{c=1}^n [-\beta m F_{\rm TAP}(\{m_{j,\alpha}^c\})
+\lambda_X X(\{m_{j,\alpha}^c\})]\right\}
\nn\\
&&\hspace*{.1cm}=
-\beta^2 m\delta_{ab}\left<\psi_a{\psi_b}\right>
+\beta^2 \langle \delta m_a \delta m_b \rangle
\nn
\\
&&\hspace*{4cm}+\frac{1}{N}\sum_i\left
\langle \frac{\partial^2\lambda_X X}{\partial h^a_i\partial
h^b_i}\right\rangle.\nn
\EEA
The very last term can be rewritten
as
\BEA
&&\frac{1}{N}\sum_i \frac{\partial^2\lambda_X X}{\partial
h^a_i\partial h^b_i}
\label{app:Ward_gen2b}\\
&&= \frac{\lambda_X}{N}\left[\delta_{ab}\sum_{i,k}\frac{\partial
X}{\partial m_k^a} \frac{\partial^2 m^a_k}{(\partial
h^a_i)^2}+
\sum_{i,k,l}\chi^a_{li}\frac{\partial^2 X}{\partial
m_k^a \partial m_l^a}\chi^a_{ki}\right]\nn
\\
&&= -2\beta\delta_{ab}\frac{1}{N}
\sum_{i,k}\lambda_X\frac{\partial  X}{\partial m_k^a}
\frac{\partial m_k^a}{\partial h_i^a}
m_i^a
+O\left(\frac{1}{\lambda_X
N},\lambda_X\right)\nn
\\
&&= -2\beta^2 \delta_{ab} \frac{1}{N}\sum_i\langle m_i^a
\delta m_i^a \rangle
+O\left(\frac{1}{\lambda_X N},\lambda_X\right)\nn
\EEA

where we have used the identity
\BEA
\frac{\partial^2 m_k}{\partial h_i^2}
=\frac{\partial^2 m_i}{\partial h_i\partial h_k}
=\beta \frac{\partial (1-m_i^2)}{\partial h_k}
=-2\beta m_i \frac{\partial m_k}{\partial h_i},
\EEA
and the definition Eq.~(\ref{f:soft_mode}) for the soft mode ${\delta\vec{ m}}$.
Furthermore, a spectral decomposition
of the susceptibility matrix  $\chi$ (as in Sec. \ref{s:stab}) shows that the last term of Eq.~(\ref{app:Ward_gen2b}), $N^{-1}\Tr[\chi
\partial ^2X \chi]$, consists of a contribution
 $(\lambda_X N)^{-1}$ due to the soft
mode, and a further contribution $\sim \lambda_X$. Both are negligible, as the
limit $N\rightarrow \infty$ is taken before $\lambda_X \rightarrow 0$.

Finally, inserting Eq.~(\ref{app:Ward_gen2b}) into Eq.~(\ref{app:Ward_gen2})
we find Eq.~(\ref{f:Ward_gen2}).


\section{Effect of regularization in the presence of many states}
\label{app:regularization}

By introducing a regularizing weight factor $\exp(\lambda_X X)$ into
the cavity formalism and the explicit sum over TAP solutions we shift
the weight slightly towards stable states.  The following constructive
description of the selected set of states may be helpful to understand
this formal trick even though we do not have a rigorous proof for its
correctness.

The selected states will be slightly stable. However, under a small
rise of the temperature they would become marginal again. We assume
that the states are selected by a weight function $X$ affecting all
states in the same (self-averaging) manner (with $X$ a symmetric
function of the $m_i$, e.g., $X=\sum_i \psi(m_i)$ with arbitrary
$\psi$.)  Then most of the selected states will become marginal at the
same slightly higher temperature $T_\lambda=T+\delta
T_\lambda$. Likewise, almost all states that are marginal at
$T_\lambda$ will adiabatically evolve into typical states selected by
$\lambda X$ at the lower temperature $T$. We my thus expect the number
of selected states to be given by
$\exp[N\Sigma(T_\lambda,f_\lambda)]$, where $f_\lambda$ is
unequivocally determined by the considered free energy density $f$ at
$T$. In general, one expects the complexity to decrease proportionally
to the increment of temperature,
\BEA
\Sigma(T,f)-\Sigma(T_\lambda,f_\lambda)\sim \delta T_\lambda.  \EEA

The properties of the selected states at $T$ follow from perturbation
theory around the marginal situation at $(T_\lambda,f_\lambda)$.  One
finds that the temperature shift $\delta T_\lambda$ induces a change
of local magnetizations of order $\Delta m_i\sim (\delta
T_\lambda)^{1/2}\zeta_i$ along the soft mode $\{\zeta_i\}$, the square
root reflecting the anomalous response of the soft mode.  Accordingly,
the value of the regularizer changes by $\delta X=\nabla
X\cdot\delta\vec{m}\sim N(\delta T_\lambda)^{1/2}$, and the soft mode
acquires a finite susceptibility $\chi^{-1}_{\rm
soft}=\sum_{ij}\zeta_i \chi^{-1}_{ij}\zeta_j \sim (\delta
T_\lambda)^{1/2}$.

The weighting procedure favors the states optimizing
$\exp[N\Sigma(T_\lambda,f_\lambda)-\lambda \delta X]$, which yields
$(\delta T_\lambda)^{1/2}\sim \lambda$, very similarly to the
mechanism in the toy model of Sec.~\ref{s:toy}. Consequently, the
susceptibility along the soft mode scales as
\BEA
\chi_{\rm soft}\sim \frac{1}{(\delta T_\lambda)^{1/2}}\sim \frac{1}{\lambda}.
\EEA
Note that all proportionality constants in the above arguments are
self-averaging constants. This implies in particular that $g\equiv
(\chi_{\rm soft} \lambda)^{-1}$ is independent of the specific
metastable state.



\begin{thebibliography}{99}
\bibitem{fn1} In the case of the SK ground state it is known that the
same methods as used for genuinely stable minima yield a correct
description.  It is unknown, however, whether this holds in general
when a multitude of soft modes is present, see also the discussion in
Sec.~\ref{ss:scenarios}.
\bibitem{fn2} However, we note that the case of the random field Ising
model is actually more subtle. A thorough analysis shows that the
description of the leading Griffith singularities requires a further
generalization of the two-group Ansatz.~\cite{MullerSilva06}
\bibitem{fn4}
The order parameters introduced
in Ref. \onlinecite{Annibale} are related to our notation by $W_{ab}\equiv
\langle x_a m_b \rangle =A_{ab}+\delta_{ab}(1-Q_{aa})$ and
$\lambda_{ab}=\beta^2/2(C_{ab}+2mA_{ab}+m^2Q_{ab})$.
\bibitem{fn5}
For every finite $\epsilon$ there are $O(N)$ eigenvalues $\lambda<\epsilon$.
\bibitem{fn6}
{The finite size scaling
 of the soft eigenvalue of the Hessian is believed to behave as $\lambda_{\rm soft}\sim N^{-1/2}$.
 The contribution of the marginal mode to the spin glass susceptibility,
$(N \lambda_{\rm soft}^2)^{-1}$, thus tends to a non-divergent constant.}
\end{thebibliography}
\end{document}